\documentclass[sigconf]{acmart}

\AtBeginDocument{%
  }

\setcopyright{acmlicensed}
\copyrightyear{2025}
\acmYear{2025}
\setcopyright{cc}
\setcctype{by}
\acmConference[CHI '25]{CHI Conference on Human Factors in Computing Systems}{April 26-May 1, 2025}{Yokohama, Japan}
\acmBooktitle{CHI Conference on Human Factors in Computing Systems (CHI '25), April 26-May 1, 2025, Yokohama, Japan}\acmDOI{10.1145/3706598.3714016}
\acmISBN{979-8-4007-1394-1/25/04}

\usepackage{enumerate}
\usepackage{multirow}
\usepackage{array} 
\usepackage{listings}
\usepackage{graphicx}

\newcommand{\red}{\textcolor[rgb]{0,0,0}}
\newcommand{\blue}{\textcolor[rgb]{0,0,0}}

\begin{document}
\title{\red{Understanding and Supporting Formal Email Exchange\\by Answering AI-Generated Questions}}

\author{Yusuke Miura}
\email{miura.yusuke@toki.waseda.jp}
\orcid{0000-0003-1204-6623}
\affiliation{%
  \institution{Waseda University}
  \city{Tokyo}
  \country{Japan}
}

\author{Chi-Lan Yang}
\email{chilan.yang@cyber.t.u-tokyo.ac.jp}
\orcid{0000-0003-0603-2807}
\affiliation{%
  \institution{The University of Tokyo}
  \city{Tokyo}
  \country{Japan}
}

\author{Masaki Kuribayashi}
\email{rugbykuribayashi@waseda.jp}
\orcid{0000-0001-8412-223X}
\affiliation{%
  \institution{Waseda University}
  \city{Tokyo}
  \country{Japan}
}

\author{Keigo Matsumoto}
\email{matsumoto@cyber.t.u-tokyo.ac.jp}
\orcid{0000-0002-0038-0678}
\affiliation{%
  \institution{The University of Tokyo}
  \city{Tokyo}
  \country{Japan}
}

\author{Hideaki Kuzuoka}
\email{kuzuoka@cyber.t.u-tokyo.ac.jp}
\orcid{0000-0003-1252-7814}
\affiliation{%
  \institution{The University of Tokyo}
  \city{Tokyo}
  \country{Japan}
}

\author{Shigeo Morishima}
\email{shigeo@waseda.jp}
\orcid{0000-0001-8859-6539}
\affiliation{%
  \institution{Waseda Research Institute for Science and Engineering}
  \city{Tokyo}
  \country{Japan}
}

\renewcommand{\shortauthors}{Miura et al.}

\begin{abstract}
\red{Replying to formal emails is time-consuming and cognitively demanding, as it requires crafting polite phrasing and providing an adequate response to the sender's demands.}
Although systems with Large Language Models (LLMs) were designed to simplify the email replying process, users still need to provide detailed prompts to obtain the expected output.
Therefore, we proposed and evaluated an \red{LLM-powered question-and-answer (QA)-based approach} for users to reply to emails by answering a set of simple and short questions generated from the incoming email.
We developed a prototype system, \textit{ResQ}, and conducted controlled and field experiments with 12 and \red{8} participants.
Our results demonstrated that \red{the QA-based approach} improves the efficiency of replying to emails and reduces workload while maintaining email quality, compared to a conventional prompt-based approach that requires users to craft appropriate prompts to obtain email drafts.
We discuss how \red{the QA-based approach} influences the email reply process and interpersonal relationship dynamics, as well as the opportunities and challenges associated with using a QA-based approach in AI-mediated communication.
\end{abstract}


\keywords{AI-Mediated Communication, Large Language Models, Email}

\begin{teaserfigure}
  \includegraphics[width=\textwidth]{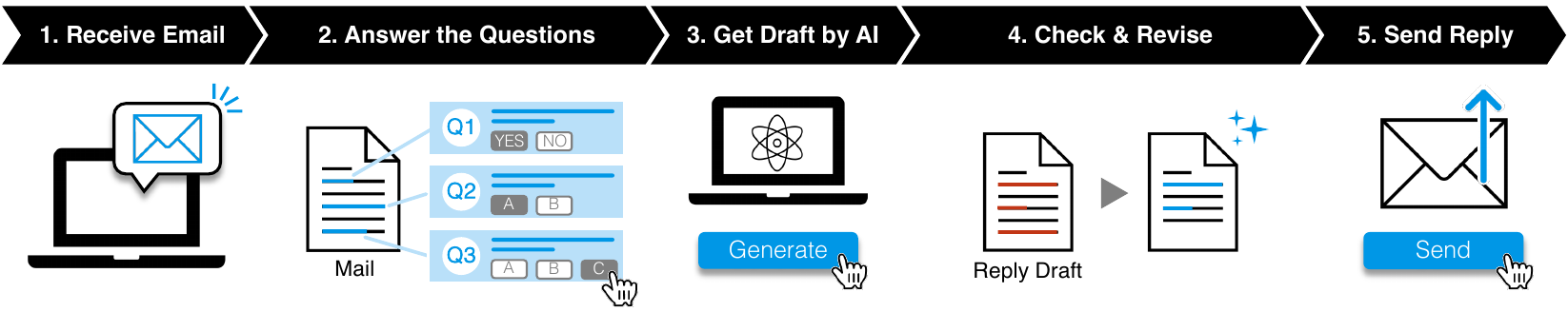}
  \caption{In our system, (1) users receive an email, (2) communicate their intentions by answering AI-generated questions, (3) receive an AI-generated draft, (4) make any necessary revisions, and finally (5) send the reply.  This process allows users to craft responses efficiently, reducing their overall workload.}
  \Description{This figure illustrates the five stages of using our proposed system. First, the user receives an email on their computer, marking the beginning of the workflow. In the second stage, the user answers a set of predefined questions related to the email, which can include options like "YES" or "NO" as well as multiple-choice selections (A, B, C). Based on the user's responses, the AI then generates a draft of the email reply in the third stage. In the fourth stage, the user reviews the AI-generated draft, checking for errors or making revisions as needed. Finally, in the fifth stage, the user sends the revised email reply, completing the process.}
  \label{fig_teaser}
\end{teaserfigure}

\maketitle
\section{Introduction}
Email is a common tool for users to share information~\cite{Nelson2011Mail2Tag, whittaker2006email} and manage tasks~\cite{bellotti2005quality}. 
It has been found that users spend an average of 28\% of their workweek reading and replying to emails~\cite{McKinseySocialEconomy}.
However, for many, checking and responding to emails is time-consuming and cognitively demanding~\cite{mark2016email}. 
Responding to emails, especially in cultures that value courteous email exchanges, requires users to understand the sender's requests and compose polite messages that reflect the sender's intentions. 
This involves considering various elements such as tone, style, diction, and structure~\cite{Chen2015Chinese}.
Putting effort into constructing courteous emails and responding promptly is important because the lack of these elements can result in a negative perception by recipients~\cite{Resendes2012Send, vignovic2010computer, yoram2011online}, potentially harming trust and damaging relationships in formal communication settings. 
\blue{In this paper, we define formal email exchange as a type of communication where the email structure is typically clear and organized, employing polite and respectful language with a specific focus. 
Examples of formal email exchange include office-related communication, research collaboration in academic institutions, and interactions with external organizations.}

Various approaches have been proposed to reduce the workload for replying to emails.
These include tools that aid in deciding whether to respond~\cite{dredze2008intelligent, di2016youvegotmail}, how to respond~\cite{vaish2017crowdtone, Respondable}, what to respond~\cite{Kannan2016Smart, Al-Alwani2013novel, Moravapalle2017DejaVuAC, Naeem2018A}, or reminding to respond~\cite{Dugan2017RemindMe:, Won2009Designing}.
With the recent advancement of generative artificial intelligence (AI), particularly large language models (LLMs), an increasing number of AI-mediated communication (AIMC) tools~\cite{ChatGPT, Grammarly, MicrosoftCopilot, bastola2024llmbasedsmartreplylsr, Claude, Foodman2022LaMPost, Fu2023Comparing, Chen2019Gmail} have been proposed.
For example, by inputting the content of an email into an AI chatbot, like ChatGPT~\cite{ChatGPT} or Claude~\cite{Claude}, along with an instruction for the model (``prompt''), these tools can generate reply drafts. 
This prompt-based response-generation approach has been shown to reduce users' overall workload and improve productivity~\cite{bastola2024llmbasedsmartreplylsr}.
While AIMC tools offer advantages, users must carefully craft prompts to achieve the desired content, tone, and style in emails~\cite{Zhou2024GlassMail}.
If the expected output is not obtained, users need to create prompts repeatedly~\cite{fu2024text}, which adds extra workload.

\red{To address this issue, we propose a QA-based approach in which the system analyzes incoming emails and generates questions that invite users to respond to efficiently create the desired draft.
In this paper, we define and implement a QA-based approach by generating questions based on the text of the email. 
Then, the system creates an email draft based on the users' answers to these questions.  
This QA-based approach was motivated by prior studies suggesting that answering structured questions helps users articulate their needs more effectively~\cite{kim2024aineedsplanner, jannach2021survey, cao2023comprehensive}. 
This approach aims to lower the cognitive burden of drafting replies, help users quickly understand the sender’s requests, and reduce the need of cumbersome prompting by breaking down the prompt-generation process into smaller, more manageable QA steps.}

Thus, this paper aims to comprehensively investigate the effect and effectiveness of the QA-based approach using \red{our prototype system, \textit{ResQ}, which generates questions and options using LLMs (Fig.~\ref{fig_teaser}).}
We also think it is important to investigate the impact of our tool on users' psychological aspects.
Therefore, we also investigate the users' sense of agency, control over the text, and perceived psychological distance, as extensive AI mediation may negatively impact these aspects~\cite{Fu2023Comparing, mieczkowski2022examining, Draxler2024The, Buschek2021The}, potentially leading to underutilization of the system.

We conducted a controlled experiment (N=12) and a field study (N=\red{8}).
We found that compared to a conventional prompt-based approach where users must consider appropriate prompts to obtain email drafts, the efficiency of replying to emails improved, and the overall workload was reduced, all while maintaining the quality of the replies.
Additionally, though ResQ lessens users’ sense of agency and control, the interview results in the field study suggest it could reduce the psychological distance from their counterparts by promoting perceptions of enhanced communication quality and quantity.

The contributions of this research are twofold. 
First, the results of two studies show how the QA-based approach\footnote{The proposed system, ResQ, will be released as an open-source Chrome extension. The source code will be publicly available at the following link: \url{https://github.com/miulab7/ResQ}.} 
affects users’ writing processes, the quality of the composed emails, and their relationships with recipients. 
Second, based on the results of these studies, we provide insights regarding both the opportunities and challenges of introducing a QA-based approach in email communication.

\section{Related Work}
\red{Here, we first describe existing AIMC systems designed to support email writing and highlight their limitations.
Second, we review QA-based approaches in goal-oriented tasks and their potential to assist with composing email replies. 
Finally, we examine the psychological and relational impacts of AIMC tools, focusing on the dynamics of recipient-sender relationships and the sender's self-perception.}

\subsection{AI-Mediated Communication Systems for Supporting Writing Emails}
Various machine learning-based approaches have been employed to enhance the productivity of writers.
Earlier writing assistants had modest AI intervention, primarily providing short or single-word suggestions~\cite{hohenstein2018AI-Supported, dunlop2012multidimensional, fowler2015effects, quinn2016cost} and basic grammar correction~\cite{Grammarly}.
With the advancement of LLMs~\cite{chang2024survey}, which allows users to obtain the long and natural-from text output by manipulating prompts, AIMC tools could enhance user input efficiency by suggesting more useful long-form text~\cite{Fu2023Comparing, dhillon2024shaping}.
Several AIMC tools allow users to select preferred tone, style, length~\cite{Grammarly, Foodman2022LaMPost, fu2024text, bastola2024llmbasedsmartreplylsr}, as well as specific content options~\cite{Grammarly, bastola2024llmbasedsmartreplylsr}, such as \textit{``decline politely''} or \textit{``ask a follow-up question''}, without manual crafting of prompt.
However, as these tools only offer simple suggestions, in situations where complex and polite replies are necessary (\textit{e.g.,} exchanging workplace emails with colleagues), users still need to carefully craft their prompts, which can impose a high workload.
Crafting effective prompts for replying to emails requires prompt engineering skills and can be a challenging task~\cite{Zhou2024GlassMail}.
When the initial output does not meet expectations, users may need to create prompts multiple times~\cite{fu2024text}, resulting in negative user satisfaction and task engagement.
\red{To address this, we propose replacing open-ended prompt creation with an LLM-powered QA-based approach, where the system leads the user through a structured question-and-answer process, effectively performing prompt engineering behind the scenes.
This method aims to reduce users’ cognitive load and reliance on prompt expertise while maintaining the quality and personalization of the email reply.}

\subsection{QA-based Approaches in Goal-Oriented Tasks}
\label{sec:2.2}
\red{Answering questions is one of the effective approaches for users to clarify their needs~\cite{kim2024aineedsplanner,aiguidebook}. 
For example, Kim~\textit{et al.}~\cite{kim2024aineedsplanner} designed a workbook using questions that guide users to organize their thoughts for developing AI projects.
In particular, answering questions has demonstrated its effectiveness in helping users get engaged in goal-oriented tasks, such as writing tasks.} 
Specifically, \red{asking questions} may be effective in extracting people's intent and aligning LLM outputs more closely with users' expectations~\cite{cao2023comprehensive}.
For instance, in conversational recommender systems~\cite{jannach2021survey}, chatbot questions are helpful in providing item suggestions tailored to users' preferences.
AI-driven questioning, inspired by Socratic methods, has been shown to promote critical thinking and improve users' ability to identify logical fallacies~\cite{danry2023dont}.
Moreover, \red{asking questions} may also facilitate task initiation, which can be particularly beneficial for individuals with a tendency to procrastinate, as they often delay responding to work-related emails~\cite{shirren2011decisional}.
This idea is derived from Fogg’s behavior model~\cite{fogg2009behavior}, which indicates that behavior change occurs when motivation, trigger, and ability converge.
Since AI outputs can capture users’ interest~\cite{brandtzaeg2017why, ling2021factors}, they may stimulate curiosity, increase motivation~\cite{berlyne1960conflict}, and serve as an effective trigger for task initiation.

\red{Applying these insights to email communication, we anticipate that a QA-based approach will not only simplify prompt creation for LLMs but also act as a cognitive scaffold for understanding incoming messages.}
\red{By breaking down the content of the received email into manageable questions, the system can highlight key information and intentions from the sender. 
This reduces the cognitive load associated with interpreting lengthy or ambiguous emails, enabling users to respond more efficiently and effectively.}
Thus, our goal is to investigate how a QA-based approach during email replies affects people's efficiency, cognitive load, task satisfaction, difficulty in initiating action, and the quality of the emails.




\subsection{The Psychological and Relational Impact of AIMC Tools}
Previous research has extensively discussed the impact of AIMC tools on the psychological aspects of both recipients and senders, as well as on their relationships. 
These effects can be broadly categorized into two areas:
(1) The impact on the recipient and sender relationships
(2) The impact on the sender’s self-perception.

\subsubsection{Impact on Recipient and Sender Relationships}
Emails can influence recipient-sender relationships through their content, speed, and context. 

Regarding the \textit{content} of the email, Robertson \textit{et al.}~\cite{Robertson2021ICant} identified three key elements of email content that, when missing, negatively impact the sender’s perception.
The first element, structural features, refers to whether the email includes structural components such as greetings, signatures, and closings. 
The second element, personal authenticity, measures the extent to which the suggested email aligns with the user’s personal tone. 
The third element, semantic and tone coherence, concerns whether the proposed email reflects the user’s intent and the broader context of the communication.
Failure to meet these criteria can lead to confusion on the recipient’s part and might reveal or raise suspicions about the use of AI, ultimately damaging the sender’s impression~\cite{hohenstein2023artificial, Jakesch2019AI-Mediated, Liu2022Will}. 
Therefore, we see the QA-based approach has the potential to address these issues by preserving the three key components of email while reflecting the user’s intent.

The \textit{speed} of email responses also influences receivers' impressions of the sender. 
Kalman and Rafaeli~\cite{yoram2011online} found that responding to business emails more quickly led to better evaluations, including increased credibility and attractiveness.
However, many people (\textit{e.g.}, administrative staff and information workers) have to manage large volumes of emails on a daily basis ~\cite{McKinseySocialEconomy}. 
As the number of incoming messages increases, response rates decline, and the content of email replies becomes shorter~\cite{kooti2015evolution}, which can potentially influence the impression formation between senders and receivers.
Our hypothesis is that a QA-based approach potentially enhances the speed of reply by lowering the barrier to task initiation.

The importance of these elements depends on \textit{context}. 
Relationships are influenced by factors such as social hierarchy, vested interests, and intimacy. 
Research indicates that in hierarchical or formal settings, both content and speed significantly affect the sender’s impression~\cite{francis2015influence, stephens2011organizational}. 
Consequently, the utility of AIMC tools also varies with social context~\cite{fu2024text, Robertson2021ICant}. 
Fu~\textit{et al.}~\cite{fu2024text} found that satisfaction with AI-generated suggestions depends on communication stakes (high vs. low) and relationship dynamics (formal vs. informal). 
They further suggested that AIMC tools are especially useful in formal settings, where established norms prevail, while their necessity decreases in informal contexts.
Thus, we mainly investigate our research questions in the formal context, where AIMC tools are known to be useful.

\subsubsection{Impact on Sender's Self-Perception}
Previous research indicates that significant AI intervention, with minimal operator input, tends to reduce people's sense of agency~\cite{Fu2023Comparing, mieczkowski2022examining} and control~\cite{Draxler2024The, Buschek2021The, kobiella2024if}. 
The agency is often characterized by action initiation~\cite{moore2012sense} and determination~\cite{bandura2001social}, both by the user. 
Sankaran~\textit{et al.}\cite{Sankaran2021Exploring} identified factors critical to maintaining people's agency, such as considering user preferences and allowing decision-making.
For instance, tasks like creating prompts and revising AI output have been found to foster a sense of accomplishment in users~\cite{kobiella2024if}.
However, in a QA-based approach, it is still unclear whether users have an increased or decreased sense of agency and control.
Thus, this study examines how the QA-based approach affects users’ sense of agency and control and how they perceive the trade-off between these feelings and the system’s benefits to better understand the approach’s advantages and risks.
\section{Research Questions and Hypotheses}
This paper aims to explore the effectiveness and potential risks of a QA-based response-writing support method by addressing the following three research questions:

\begin{enumerate}[RQ1:]
    \item How does a QA-based response-writing support approach affect users’ email-replying process?
    \item How does a QA-based response-writing support approach affect the quality of the email response?
    \item How does a QA-based response-writing support approach affect the perceived relationship between email sender and recipient?
\end{enumerate}

To answer the three research questions, we formed three sets of hypotheses.
The first set of hypotheses investigates the impact of a QA-based system on users' email-replying process.
AI-powered text generation reduces user input, saves time, and enhances efficiency~\cite{bastola2024llmbasedsmartreplylsr}. 
It also helps users quickly grasp email content with less cognitive effort, particularly through text summarization and list formatting, which enhances productivity~\cite{tarnpradab2017toward, nandhini2013use, modaresi2017commercial, daniel1998influence}. 
This suggests that presenting questions in a list format could streamline email responses, reducing the need for detailed prompts. 
Based on these insights, we propose the following hypotheses:

\begin{enumerate}[\textrm{H1-}a:]
    \item QA-based system enhances users’ email replying efficiency.
    \item QA-based system reduces users’ cognitive load while replying to email.
\end{enumerate}

As a result, we expect users’ perceived work efficiency to improve.
Furthermore, since the QA-based system suggests appropriate language and helps create responses that align with the recipient’s needs, we anticipate that users’ satisfaction with their email replies will increase.
Thus, we propose the following hypothesis:
\begin{enumerate}[\textrm{H1-}c:]
    \item QA-based system enhances users' satisfaction with completing email response tasks, thereby being favorably received by users.
\end{enumerate}

Moreover, as described above, reducing users’ burden and improving their satisfaction may enhance their confidence in their tasks, which could lower their hesitation to begin working~\cite{schouwenburg1992procrastinators}.
Additionally, AI outputs that engage users’ curiosity may help trigger task initiation~\cite{brandtzaeg2017why, ling2021factors}.
Thus, we propose the following hypothesis:
\begin{enumerate}[\textrm{H1-}d:]
    \item QA-based system lowers the barriers to initiating email response tasks.
\end{enumerate}

According to previous research, there is a trade-off between the degree of AI intervention and the sense of agency and control, with higher levels of AI involvement shown to diminish these perceptions~\cite{Fu2023Comparing, Draxler2024The}.
Given that our QA-based approach also involves AI intervention during the phase where users create prompts for the LLM, the following hypothesis can be derived:
\begin{enumerate}[\textrm{H1-}e:]
    \item QA-based system diminishes users’ sense of agency and reduces their sense of control of the content.
\end{enumerate}

The second hypothesis concerns the quality of email responses.
AI support can be helpful in ensuring appropriate language use and grammar~\cite{fu2024text}. 
Furthermore, the QA-based approach is expected to assist users in correctly understanding the intent and demands of received emails and in verifying whether their responses meet these requirements.
Based on this, we propose the following hypothesis:
\begin{enumerate}[\textrm{H2}:]
    \item QA-based system enhances the perceived quality of the email response.
\end{enumerate}

The third set of hypotheses investigates the perceived relationship between email sender and recipient.
When users use the QA-based approach, it is expected that their communication partners will receive high-quality messages more quickly. 
Thus, the following hypothesis is derived.
\begin{enumerate}[\textrm{H3-}a:]
    \item QA-based system makes a good impression on the user's communication partner.
\end{enumerate}

On the other hand, when users create messages using the AIMC tool, they may feel a sense of discomfort with the message and guilt for not having fully composed it themselves~\cite{fu2024text}. 
We hypothesized that a QA-based approach would further intensify this discomfort by reducing the user’s sense of agency and control more than previous approaches.
\begin{enumerate}[\textrm{H3-}b:]
    \item QA-based system enlarges the psychological distance that users perceive toward their communication partners.
\end{enumerate}

\section{Proposed LLM-Powered QA-Based Approach: ResQ}
\label{sec:Proposed_Approach}
This section describes the proposed approach, ResQ, for supporting email response tasks. 
\begin{figure*}[t]
\centering
\includegraphics[width=\textwidth]{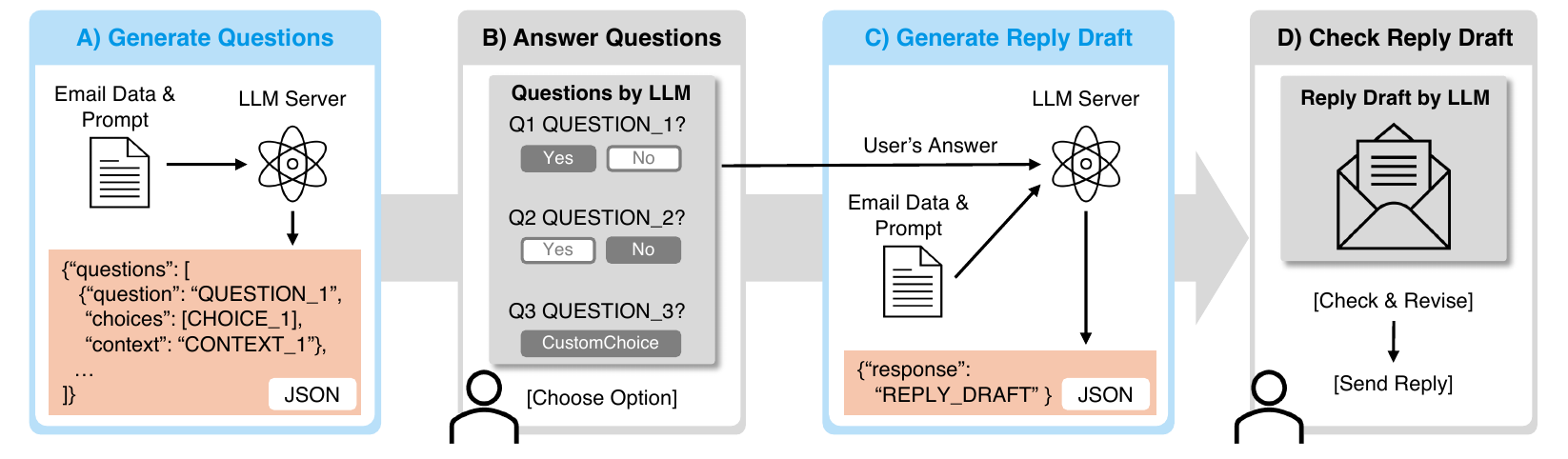}
\caption{The overview of the process of creating a reply message using ResQ. A) The LLM first generates multiple-choice questions in JSON format. B) Users select their desired responses to their counterparts. C) The LLM then generates a reply draft in JSON format based on the users' selections. D) Finally, users review and edit the LLM-generated draft before sending the reply.}
\label{fig_overview_of_process}
\Description{This figure illustrates the process of generating an email reply using a large language model (LLM) across four stages. In Stage A (Generate Questions), the system takes the email data and a prompt, which are then sent to the LLM server. The server processes this information and generates a set of questions to clarify the content of the reply. These questions, along with possible answer choices and context, are returned in JSON format. In Stage B (Answer Questions), the user is presented with the questions generated by the LLM. These questions may include simple "Yes" or "No" options or multiple-choice selections. The user answers the questions by choosing the appropriate options or providing custom responses. In Stage C (Generate Reply Draft), the user’s answers, along with the original email data and prompt, are sent back to the LLM server. Based on this input, the server generates a draft of the email reply, which is also returned in JSON format. In Stage D (Check Reply Draft), the user reviews the draft generated by the LLM. After checking the content and making any necessary revisions, the user finalizes and sends the email reply.}
\end{figure*}
\begin{figure*}[t]
\centering
\includegraphics[width=\textwidth]{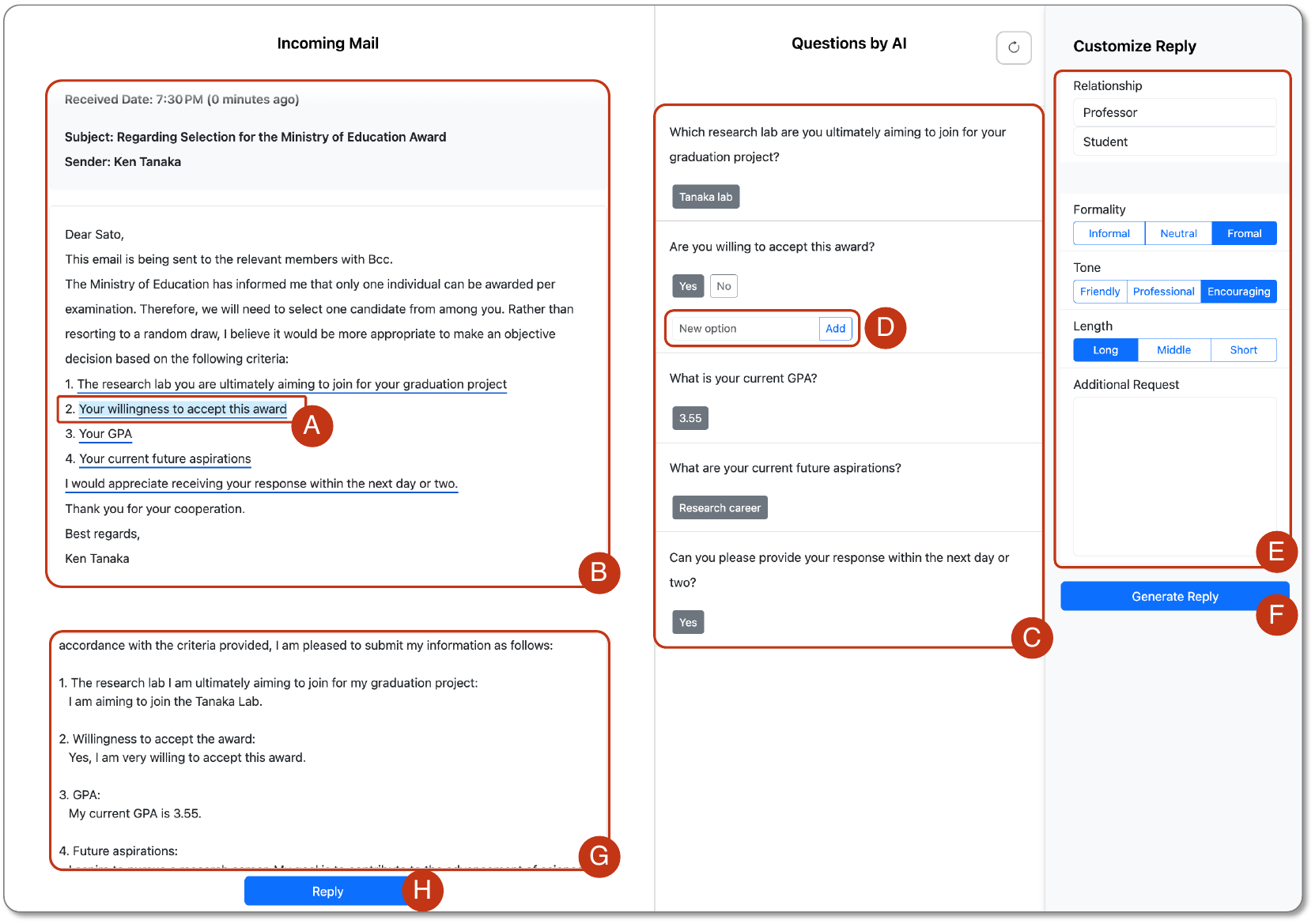}
\caption{Interface of ResQ. On the left, the content of the email is displayed, with an editor and a ``Reply'' button below for sending a reply. In the center, questions and options for users are shown, allowing the creation of custom options if needed. Additionally, the section of the email corresponding to the selected question is highlighted. On the right, fields are provided to customize the reply generated by the LLM, including options to specify the relationship with the counterpart and buttons to choose the formality, tone, and length of the email. A free-text input field and a "Generate Reply" button are also below.
}
\label{fig_interface}
\Description{This figure illustrates the interface of the ResQ system, which is divided into three main sections. On the left side (A, B), the content of the incoming email is displayed. The email includes important information, such as the sender, subject, and body text. Key sections of the email are highlighted based on the questions generated by the system, helping the user focus on relevant points. Below the email, there is an editor where the user can compose their reply, with a "Reply" button (H) available to send the response once it's ready. In the center section (C, D), the system displays questions generated by the AI, which are intended to assist the user in composing their reply. These questions correspond to specific parts of the email, and as the user answers them, the relevant section in the email is highlighted (A). Users can also customize responses by adding new options if needed (D). user can specify their relationship with the email recipient (e.g., professor or student), adjust the formality and tone of the response, and select the desired length of the reply. An additional free-text input field is available for further customization requests (E). Once all preferences are set, the user can click the "Generate Reply" button (F) to produce a draft response based on their inputs.}
\end{figure*}
Fig.~\ref{fig_overview_of_process} illustrates the overview of how a reply message is created using ResQ.
Fig.~\ref{fig_interface} shows the actual interface of ResQ.
The following sections describe the specific functions involved in each step of this process.

\paragraph{\textbf{A: Generate Questions}}
\label{sec:generate_questions}
When a user first activates ResQ, the system uses an LLM (in this study, GPT-4o~\cite{GPT4o}) to generate multiple-choice questions (Fig.~\ref{fig_interface}-C). 
The LLM extracts all parts of the email that require a reply, generates corresponding questions and presents possible response options.
Additionally, if a user clicks on any generated question, the relevant part of the email is highlighted (Fig.~\ref{fig_interface}-A).

\red{Following the approach described in~\cite{bsharat2023principled}, we designed a structured prompt that guides the LLM to determine how many questions are necessary and sufficient to cover all requirements in the incoming email without omission. 
Instead of pre-specifying a fixed number of questions, the prompt instructs the model to produce an ``appropriate'' number of questions, where ``appropriate'' is defined as the minimal set of questions needed to address all points raised by the sender while avoiding redundant or irrelevant inquiries.
To ensure that the questions were generated systematically rather than randomly, we provided explicit criteria within the prompt. 
These criteria included referencing the sender's intent, quoting relevant portions of the original email verbatim, and offering multiple-choice options where applicable. 
We also provided concrete examples within the prompt to illustrate the desired format and style of the generated questions and corresponding answer choices. 
By doing so, we ensured that the LLM's output was both well-grounded and easy for the recipient to answer.
The detailed prompt used to guide the LLM in this process is shown in the appendix.}


\paragraph{\textbf{B: Answer Questions}}
Next, users view the incoming email (Fig.~\ref{fig_interface}-B) alongside the generated questions (Fig.~\ref{fig_interface}-C) and options and proceed to answer them. 
In anticipation of situations where none of the provided options are useful, we enabled users to add their own options (Fig.~\ref{fig_interface}-D). 
Additionally, to help the LLM better understand the context of the email, we introduced a box where users can specify the relationship between the sender and the recipient (Fig.~\ref{fig_interface}-E, top). 
Furthermore, following previous research~\cite{fu2024text}, we provided users with controls to adjust the tone, style, and length of the reply to match their preferences better, thereby giving them more flexibility in customizing the AI-generated response (Fig.~\ref{fig_interface}-E, middle). 
A free-text field was also included to allow users to make other specific AI requests (Fig.~\ref{fig_interface}-E, bottom). 
After completing these steps, users can click the ``Generate Reply'' button (Fig.~\ref{fig_interface}-F).

\paragraph{\textbf{C: Generate Reply Draft}}
When the user clicks the ``Generate Reply'' button, ResQ detects the action and uses the LLM to generate a reply draft.
The prompt used for this function is shown in the appendix.

\paragraph{\textbf{D: Review Reply Draft}}
Once the draft reply is generated, users can review the draft in detail (Fig.~\ref{fig_interface}-G).
Moreover, if users find that extensive revisions are needed or if they want to explore alternative phrasing, they have the option to request the AI to regenerate a new draft based on updated input or preferences.
After completing these steps, users can click the ``Reply'' button (Fig.~\ref{fig_interface}-H).
\section{Method of Study 1}
\red{To test our hypotheses and answer three research questions, we first conducted a controlled experiment, focusing on gaining a quantitative understanding.}

\subsection{Experiment Design}
Study 1 was designed to quantitatively assess how ResQ influences the writing process (RQ1), the quality of email replies (RQ2), and the perceived relationships with others (RQ3) compared to scenarios without AI intervention and when using traditional AIMC tools. 
\red{The experiment targeted Japanese participants and was conducted entirely in Japanese.}
This experimental context was designed as communication in formal settings, such as \red{office-related communication, research collaboration in academic institutions, and interactions with external organizations}. 
\red{It focused on time-consuming emails that were characterized as lengthy, containing multiple requests, or requiring detailed and polite responses. 
Simple and straightforward emails, such as those that can be answered with a single word or phrase (\textit{e.g.}, ``Understood''), were excluded from the scope.}

Participants were assigned the role of message recipients and required to craft replies based on the scenarios and supplementary information provided. 
\red{The messages used in the experiment were collected from 10 volunteers who provided real emails they had received in formal communication contexts.}
\red{These volunteers included office workers, graduate students, and teaching staff, all of whom were Japanese and engaged in email communication regularly (at least once per month).}
To ensure anonymity, identifying details were removed during the preparation process. 
\red{Based on the design principles of ResQ, we excluded extremely short emails and emails that could be replied to with a single word from the selection process.}
Each scenario included details about the sender, the recipient, and the context in which the message was received. 
Multiple scenarios were included in the experiment to minimize the influence of any single scenario and increase the variety. 
Additionally, supplementary information, such as the recipient’s schedule and potential questions, was provided to prevent excessive variability in the responses among participants. 

In total, we created twenty types of \red{emails}, with two assigned to the practice session and eighteen to the test session.
The length of the \red{emails} used in the test session averaged 404 \red{Japanese} characters, with the shortest being 135 characters and the longest being 925 characters.
\red{The scenarios covered a wide range of formal communication situations, including responding to a request for data submission in the workplace, answering a survey from a professor, asking questions based on guidance from a language school’s customer support team, and addressing a request for schedule adjustments as a part-time worker.}
\red{Additionally, these emails varied in structure, ranging from structured formats with bullet points to more non-structured, free-text formats.}
\red{The specific emails and examples of ResQ-generated questions and options used in the experiment can be referred to in the supplementary materials.}

\subsection{Experimental Conditions}
We employed a within-subject design with three conditions: QA-based, Prompt-based, and No-AI.
\red{This design was chosen to control for individual differences among participants, such as varying levels of language proficiency or familiarity with AI systems, ensuring a fair comparison across conditions.}
To illustrate each condition, consider the scenario of a participant who, as an employee of a company, is asked by their superiors to assume the role of a fixed asset committee member.

In the QA-based condition, participants created replies using the QA-based AI.
The system detected when participants navigated to the next email screen, inferred that they were initiating a reply, and then generated relevant questions.
For example, the system might ask, ``Would you be willing to take on the role of the fixed asset manager?'' ``Is there any issue with handling the annual inventory check?'' or ``Please let us know if you have any questions or concerns about the tasks.''
Participants could respond by selecting from provided options (\textit{e.g.}, ``yes,'' ``no''), adding their own options, or ignoring the questions entirely. 
After responding, they would press the ``Generate Reply'' button, which would produce an AI-generated draft in the reply box. 
Participants could then regenerate or modify the draft as needed to finalize their response.

In the Prompt-based condition, participants created replies using a prompt-based AI without the QA feature of the QA-based method.
Participants wrote prompts for the AI to generate a draft email response, which they then edited to create their replies.
For example, a participant might input a prompt such as, ``I want to convey my acceptance of the fixed asset committee role. ...''
Afterward, similar to the QA-based system, participants would press the ``Generate Reply'' button and, if necessary, either regenerate the draft or revise its content.

In the No-AI condition, participants created email replies manually without using AI assistance.

\subsection{Participants}
\red{
\begin{table*}[t]
\caption{Backgrounds of participants in Study 1, including age, job roles, email experience, frequency of email sending and AI tool usage, and use of AI for email purposes. \blue{Some fields are marked as - due to missing responses from participants.}}
\Description{The table provides an overview of participants in Study 1, detailing their demographic information, email usage habits, and AI tool usage patterns. The participants, ranging in age from 20 to 57, include university students, office workers, a part-time worker, and individuals categorized under "other" occupations. Both male and female participants are represented, reflecting a diverse group in terms of age, occupation, and technological engagement. Participant P1 is a 34-year-old male office worker with 20 years of email experience. He typically replies to 7 emails per week but rarely uses AI tools, and he never utilizes AI for email-related tasks. P2, a 22-year-old male university student, has 5 years of email experience and actively engages with emails, replying to more than 21 per week. He uses AI tools daily, with AI assisting in 50–80\% of his email-related tasks. Similarly, P3, a 22-year-old female university student, reports no specific email experience or weekly email activity but uses AI tools frequently, relying on AI for 50–80\% of her email tasks. P4, a 21-year-old female university student with 4 years of email experience, replies to 0–2 emails weekly. She uses AI tools frequently but for less than 20\% of her email-related activities. P5, a 20-year-old male university student, has no reported email experience or weekly email activity. He rarely uses AI tools and does not use them for email purposes. P6, a 38-year-old male office worker, has 3 years of email experience and replies to 0–2 emails weekly. He rarely engages with AI tools and never applies them to email tasks. P7, a 31-year-old female categorized as "unemployed," has 12 years of email experience and replies to 3–4 emails per week. She never uses AI tools, either for general purposes or for email tasks. In contrast, P8, a 25-year-old male university student with 4 years of email experience, replies to 0–2 emails per week. He uses AI tools frequently, with AI assisting in 50–80\% of his email-related activities. P9, a 39-year-old female office worker, has 20 years of email experience and replies to over 21 emails per week. She uses AI tools frequently but never applies them to email tasks. P10, a 24-year-old male university student, has no reported email experience or weekly activity. He uses AI tools daily, although only for less than 20\% of his email-related tasks. P11, a 22-year-old female university student with 4 years of email experience, replies to 0–2 emails weekly. She rarely engages with AI tools and does not use them for email tasks. Finally, P12, a 57-year-old female office worker with 20 years of email experience, replies to 0–2 emails weekly. She uses AI tools frequently but never for email-related purposes.}
\label{tab_study1_participants}
\red{
\begin{tabular}{cccccccc}
\hline
ID   & Gender & Age & Job                  &  Email Experience&Emails/Week & AI Tool Usage& AI for Email Usage    \\ \hline
P1   & M& 34  & Office Worker        &  20 years&7           & Rarely         & Never                 \\
P2   & M& 22  & Univ. Student        &  5 years&21+         & Daily              & 50–80\%\\
P3   & F& 22  & Univ. Student        &  -&-& Frequently& 50–80\%\\
P4   & F& 21  & Univ. Student        &  4 years&0–2         & Frequently& <20\%\\
P5   & M& 20  & Univ. Student        &  -&-           & Rarely         & <20\%\\
P6   & M& 38  & Office Worker        &  3 years&0–2         & Rarely         & Never                 \\
P7   & F& 31  & \blue{Unemployed}                &  12 years&3–4         & Never         & Never                 \\
P8   & M& 25  & Univ. Student        &  4 years&0–2         & Frequently& 50–80\%\\
P9   & F& 39  & Office Worker        &  20 years&21+         & Frequently& Never                 \\
P10  & M& 24  & Univ. Student        &  -&-           & Daily              & <20\%\\
P11  & F& 22  & \blue{Univ. Student}     &  4 years&0–2         & Rarely         & Never                 \\
P12  & F& 57  & Office Worker        &  20 years&0–2         & Frequently& Never                 \\ \hline
\end{tabular}
}
\end{table*}
}

Twelve participants (six males and six females, aged 20-57) were recruited via a local Japanese participant recruiting platform (see Tab.~\ref{tab_study1_participants}).
The average age of the participants was 29.6 (SD = 11.0). 
\red{The sample size $n=12$ was determined based on an a priori power analysis (effect size $f=0.4$, significance level $p=0.05$, power = 0.8, correlation among repeated measures = 0.5) as well as the previous study~\cite{Mu2024Whispering}.}
The participants were paid approximately \$21 USD for participation, and the experiment lasted around two hours.
This study was approved by the institute's ethical review board.

\subsection{Procedure}
The participants first read the study instructions and the right to participate and then consented to participate in the experiment. 
Next, they were given an explanation of the purpose of the experiment and the use of the AI systems (Prompt-based and QA-based systems). 
\red{Participants were then randomly assigned to reply to six emails per condition using a Latin square design, which counterbalanced the order of conditions and mitigated potential order effects\blue{~\footnote{\blue{We conducted analyses to examine the potential order effect. The results of this analysis are provided in the appendix.}}}.}
In each condition, participants first engaged in a practice session where they read and replied to two emails \red{to familiarize themselves with the system.}
Then, they read and replied to six emails, which were presented in a randomly assigned order \red{to further reduce any sequence-related biases.}
After replying to six emails for each condition, participants were asked to complete a questionnaire regarding their experience with the task. 
\blue{To ensure participants could manage their workload during the study, they were allowed to take a short break after completing tasks in each condition. 
}
After completing all conditions, they were asked to fill out a comparative questionnaire evaluating the three conditions. 
\red{In addition, follow-up interviews were conducted to gather deeper insights into their experiences and preferences.}
This study was conducted remotely for all participants and lasted approximately two \red{and a half} hours in total.

\subsection{\red{Evaluation Session}}
After completing the main experiment, we conducted an additional evaluation session to assess the quality of the email responses created by participants and the impressions of participants as email senders.
This session involved a group of eighteen Japanese evaluators (ten males and eight females, aged 20-57) recruited via a local participant recruiting platform~\footnote{Participants were recruited from Lancers.jp, an online freelancing platform.}. 
The average age of the evaluators was 40.6 (SD = 8.3).
\blue{The evaluators had a minimum of five years and an average of 18.7 years of experience in email-based communication. 
Furthermore, with the exception of one individual, the evaluators engaged in email-based communication at least once a month.}
Each evaluator assessed email replies written by twelve different participants for a specific scenario. 
The evaluators were paid approximately \$2.5 USD for their participation, and the evaluation session lasted around fifteen minutes.

\subsection{Measurements}
\red{We used multiple measurements to test our hypotheses. 
From participants' behavior during the email reply task, we calculated two measures: efficiency and prompt character count. 
From their post-experiment questionnaire responses, we evaluated cognitive load, difficulty in understanding email content, satisfaction with completing the task, difficulty in initiating the action for replying to emails, sense of agency, sense of control, and psychological distance between participants and their counterparts. 
Additionally, from evaluators’ questionnaire responses during the evaluation session, we assessed the perceived quality of the email and the impression of participants as email senders.}
\subsubsection{Efficiency of Replying to Emails (H1-a)}
We calculated the efficiency of replying to emails using task completion time and total character count.
The efficiency of replying to emails is defined as the amount of text contributing to the final output that can be typed per second, where a higher score indicates better task efficiency.
For task completion time, we recorded the time participants took to reply to an email, starting from when the email appeared on the screen to when the participant pressed the send button.
For total character count, we considered the text in the reply box when the participant pressed the Reply button as the final response and counted its characters.

\subsubsection{Prompt Character Counts (H1-a)}
We also calculated the average number of characters typed by participants to have the AI generate email drafts as the prompt character counts in each condition.
Under the Prompt-based condition, we measured the number of characters participants typed in the free-text field for the AI. 
Under the QA-based condition, the prompt character counts included this number plus any additional characters typed by the participants when they added their own options.

\subsubsection{Cognitive Load for Replying to Emails (H1-b)}
We used the NASA-TLX~\cite{hart1988development} questionnaire to measure cognitive load across six subscales: mental demand, physical demand, temporal demand, performance, effort, and frustration and calculated the Raw-TLX~\cite{byers1989traditional}.
Participants answered the above items using a 10-point Likert scale.
The Raw-TLX score is calculated as the simple average of six scales, where higher scores indicate a greater cognitive load.

\subsubsection{\red{Difficulty in Understanding Email Content (H1-b)}}
\red{Additionally, to assess cognitive load specifically related to understanding received emails, we used a 7-point Likert scale. 
Participants rated their agreement with the statement, ``I found it difficult to understand the sender’s intentions or requests in the email,’’ where 1 indicates strongly disagree, 4 indicates neutral, and 7 indicates strongly agree.}


\subsubsection{Satisfaction with Completing Task (H1-c)}
We evaluated participants' satisfaction with completing their task using a 7-point Likert scale, where 1 indicates strongly disagree, 4 indicates neutral, and 7 indicates strongly agree.
Specifically, the satisfaction of completing their task was evaluated based on their satisfaction with efficiency and their satisfaction with the quality of the email they created.
We asked the following questions: 
(1) I felt that I was able to create a high-quality response. 
(2) I felt that I was able to complete the response efficiently.
We averaged the scores from two items and treated them as an index of the satisfaction with completing their task.


\subsubsection{Difficulty in Initiating the Action for Replying to Emails (H1-d)}
We tested H1-d using a survey with a 7-point Likert scale (1 = strongly disagree, 4 = neutral, and 7 = strongly agree) to evaluate perceived barriers to task initiation. 
Specifically, we asked the following question: I felt a high barrier to initiating email response tasks.

\subsubsection{Sense of Agency and Control (H1-e)}
We evaluated participants' perceived sense of agency and control using a 7-point Likert scale, where 1 indicates strongly disagree, 4 indicates neutral, and 7 indicates strongly agree.
Specifically, drawing on previous research~\cite{Fu2023Comparing, Draxler2024The}, the sense of agency was evaluated by assessing whether participants felt they were the ones who wrote the responses, while the sense of control was evaluated by whether they felt they had control over the content of the responses. 

\subsubsection{Perceived Quality of the Email by Evaluators (H2)}
\label{sec:method2_H2}
The quality of each email reply was evaluated \red{by evaluators} using a 7-point Likert scale, where 1 indicates strongly disagree, 4 indicates neutral, and 7 indicates strongly agree.
It was evaluated on three aspects: politeness (whether it was politely written), readability (whether it had an easy-to-understand structure), and meeting demands (whether it appropriately addressed the recipient's demands). 
We averaged the scores from three items and treated it as an index of the perceived quality of the email.

\subsubsection{\blue{Perceived Impression of Participants by Evaluators (H3-a)}}
\red{Following a previous study~\cite{rau2009effects}, we asked the evaluators to read the email and assess their impressions of the senders (participants) based on two aspects: whether the participants were perceived as likable and whether they were perceived as kind, using a 7-point Likert scale, where 1 indicates strongly disagree, 4 indicates neutral, and 7 indicates strongly agree. 
We averaged the scores from these two items to create an index of the impression of the email sender.}


\subsubsection{Psychological Distance between Participants and Their Counterpart (H3-b)}
\begin{figure*}[t]
\centering
\includegraphics[width=\textwidth]{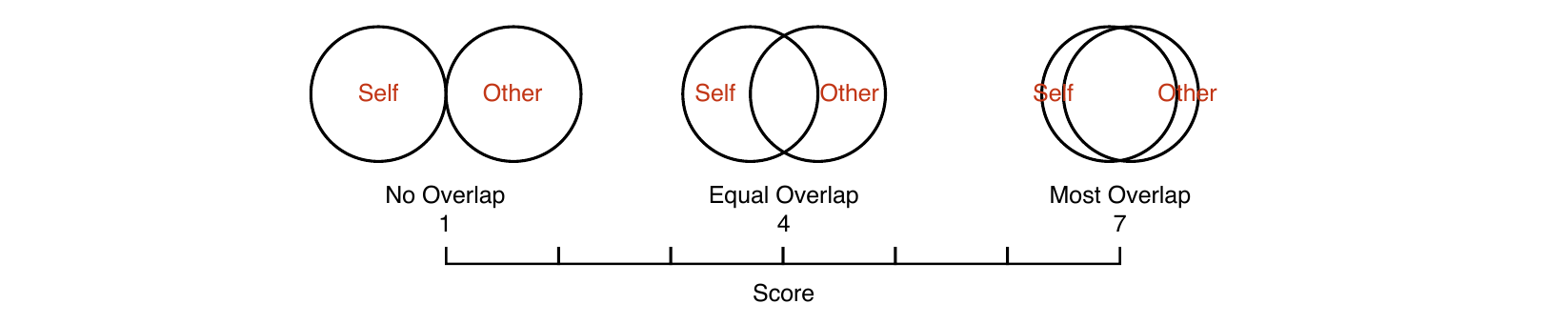}
\caption{Inclusion of Other in the Self (IOS). The diagram above the x-axis is an example of what participants were shown when responding to the questionnaire. The degree of overlap between the two circles represents the psychological distance between oneself and others.}
\label{fig_study1_IOS}
\Description{This figure represents the "Inclusion of Other in the Self (IOS)" scale, which is used to measure psychological closeness or relational intimacy. The diagram depicts two circles, labeled "Self" and "Other," with varying degrees of overlap. Participants were asked to choose the level of overlap that best represented their relationship with another person. On the far left (score 1), the circles are completely separate, indicating a significant psychological distance between the self and the other. In the middle (score 4), the circles partially overlap, suggesting a moderate level of psychological closeness. On the far right (score 7), the circles almost completely overlap, representing a very close and intimate relationship between the self and the other.}
\end{figure*}
We evaluated the perceived psychological distance using the Inclusion of Other in the Self (IOS) scale~\cite{aron1992inclusion}.
Participants choose a pair of circles from seven with different degrees of overlap (\red{see Fig.~\ref{fig_study1_IOS}}). 
1 = no overlap; 2 = little overlap; 3 = some overlap; 4 = equal overlap; 5 = strong overlap; 6 = very strong overlap; 7 = most overlap. 
The number chosen is the participants’ score.
The higher the score was, the closer participants felt they were with the email sender.
\begin{figure*}[t]
\centering
\includegraphics[width=\textwidth]{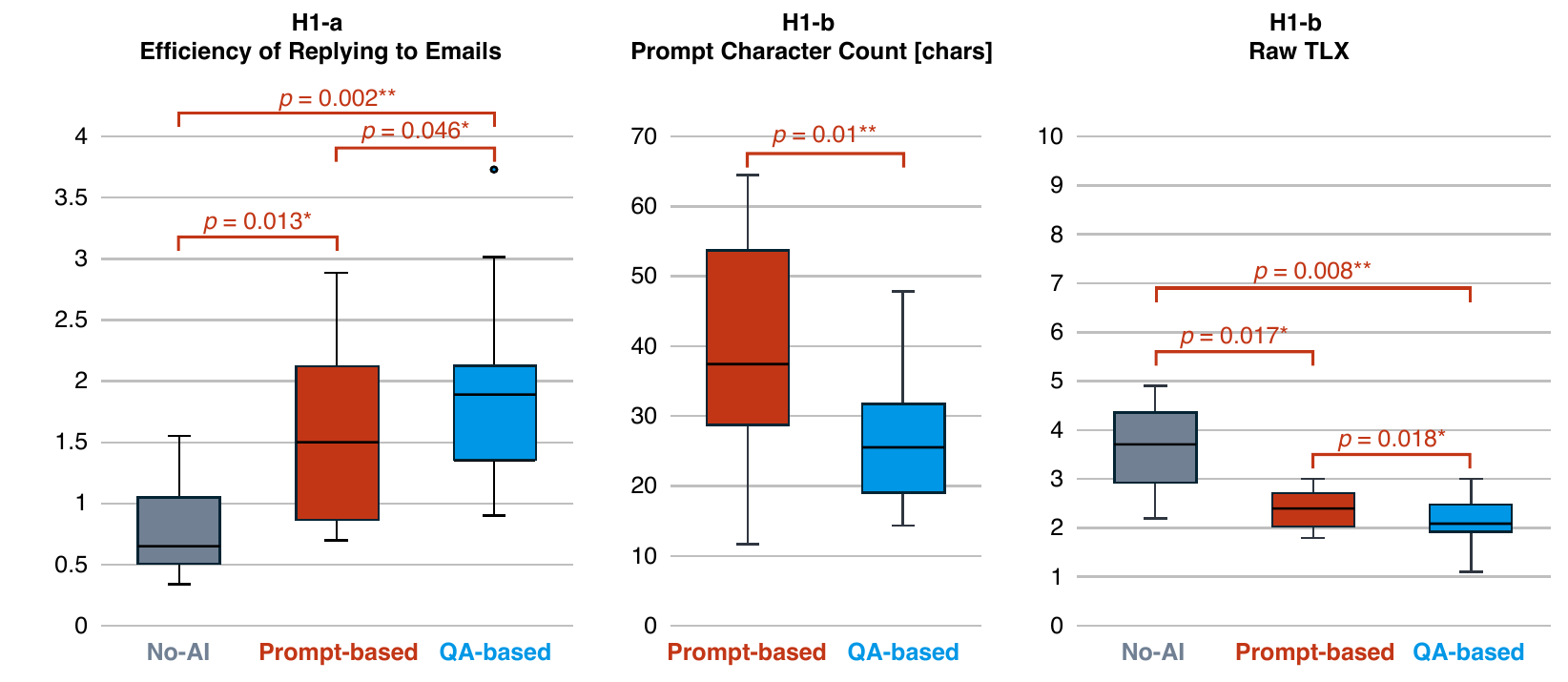}
\caption{Results of participants' efficiency and cognitive load of replying to emails. Left: Efficiency for replying to emails. Middle: Prompt character count. Right: Cognitive load for replying to emails. The significant differences between conditions were from post-hoc analysis after doing one-way repeated measure ANOVA.}
\label{fig_study1_efficiency_and_cognitiveLoad}
\Description{The figure consists of three box plots, each representing different comparisons of three experimental conditions: No-AI, Prompt-based, and QA-based. Each box plot compares a specific measure between the conditions. The p-values indicating statistical significance between different conditions are also labeled above the plots. Efficiency of Replying to Emails: Three conditions are compared: No-AI, Prompt-based, and QA-based. The No-AI group has a median of 0.65, with a first quartile at 0.51 and a third quartile at 1.05, with a minimum of 0.34 and a maximum of 1.55. The Prompt-based group has a median of 1.50, with the first quartile at 0.87 and the third quartile at 2.12, with a minimum of 0.70 and a maximum of 2.89. The QA-based group has a median of 1.89, with a first quartile at 1.35 and a third quartile at 2.12, with a minimum of 0.90 and a maximum of 3.73. P-values indicate significant differences between groups: between No-AI and Prompt-based (p = 0.013), between No-AI and QA-based (p = 0.002), and between Prompt-based and QA-based (p = 0.046). Prompt Character Count: Only two conditions are compared: Prompt-based and QA-based. The Prompt-based group has a median of 37.42 characters, with a first quartile at 28.75 and a third quartile at 53.50, with a minimum of 11.67 and a maximum of 64.50. The QA-based group has a median of 25.50 characters, with a first quartile at 19.00 and a third quartile at 31.67, with a minimum of 14.33 and a maximum of 47.83. The p-value indicates a significant difference between the two groups (p = 0.01). Raw TLX: This plot compares three conditions: No-AI, Prompt-based, and QA-based. The No-AI group has a median score of 3.70, with a first quartile of 2.93 and a third quartile at 4.35, with a minimum of 2.20 and a maximum of 4.90. The Prompt-based group shows a median of 2.40, with a first quartile at 2.03 and a third quartile at 2.70, with a minimum of 1.00 and a maximum of 4.80. The QA-based group has a median of 2.10, with a first quartile of 1.93 and a third quartile at 2.48, with a minimum of 0.70 and a maximum of 4.40. The p-values indicate significant differences between No-AI and Prompt-based (p = 0.017), between No-AI and QA-based (p = 0.008), and between Prompt-based and QA-based (p = 0.018). Each box plot represents the distribution of values for the respective metric, and the whiskers show the variability outside the upper and lower quartiles. The statistical differences (p-values) highlight where the comparisons between conditions are significant.}
\end{figure*}
\begin{figure*}[t]
\centering
\includegraphics[width=\textwidth]{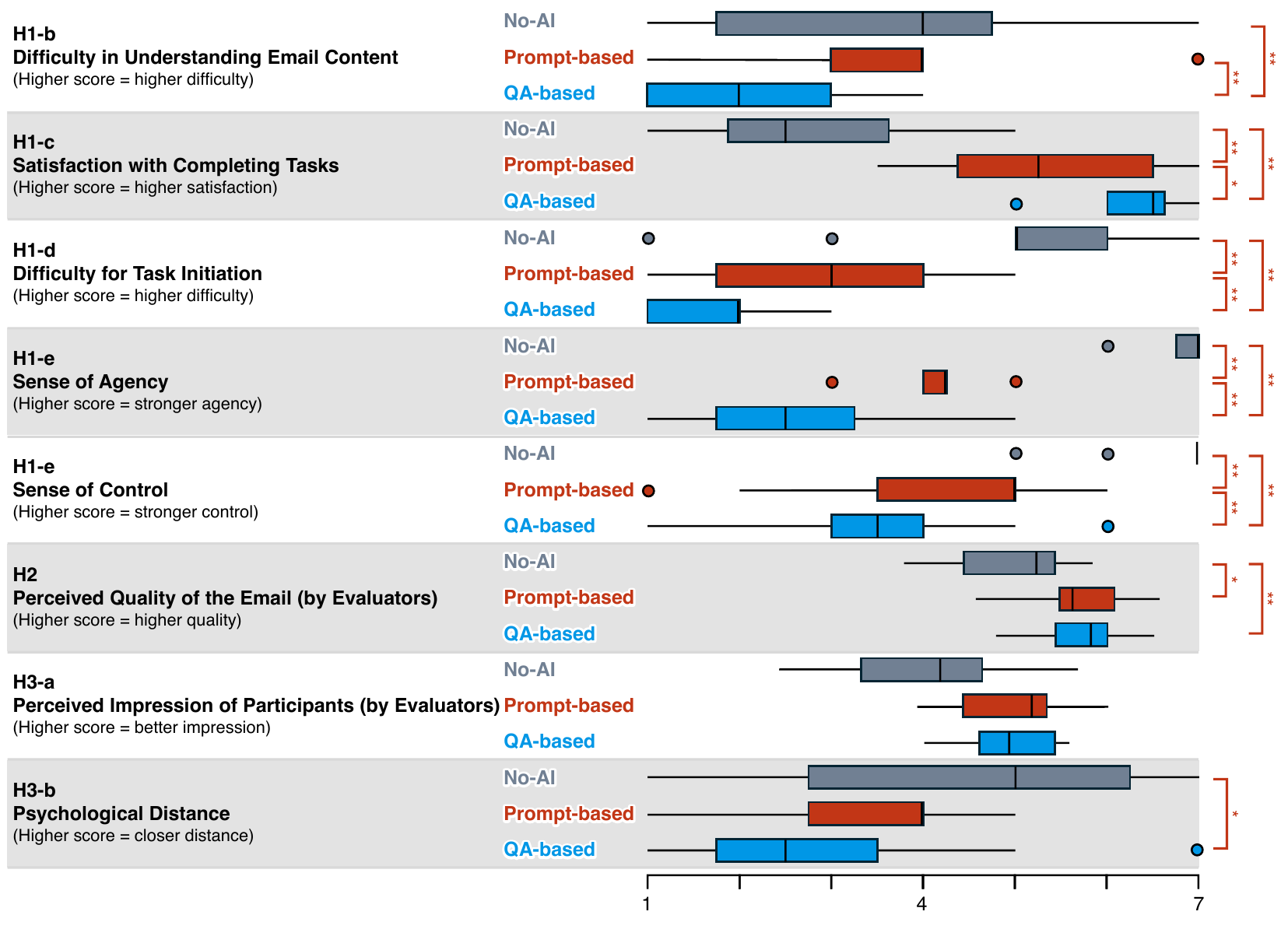}
\caption{Summary of Likert scale responses. \red{Measurements H2 and H3-a were assessed by third-party evaluators rather than the participants themselves.} The significant differences between conditions were from post-hoc analysis after one-way repeated measure ANOVA or the Friedman test (* and \red{**} indicate the significance found at levels of 0.05 and 0.01, respectively).}
\label{fig_study1_questionnaire}
\Description{The figure consists of box plots comparing three experimental conditions, No-AI, Prompt-based, and QA-based, across multiple subjective measures related to email task performance. The box plots represent the distribution of responses across these conditions, with p-values indicating statistically significant differences between them. H1-b: Difficulty in Understanding Email Content: No-AI: The median is 4, with a first quartile at 3.25 and a third quartile at 6.25, with a minimum of 1 and a maximum of 7. Prompt-based: The median is 4, with a first quartile at 4 and a third quartile at 5.25, with a minimum of 1 and a maximum of 7. QA-based: The median is 6, with a first quartile at 5 and a third quartile at 7, with a minimum of 4 and a maximum of 7. Significant differences exist between Prompt-based and QA-based (p < 0.01) and between No-AI and QA-based (p < 0.01). H1-c: Satisfaction with Completing Tasks: No-AI: The median is 2.5, with a first quartile at 1.875 and a third quartile at 3.625, with a minimum of 1 and a maximum of 5. Prompt-based: The median is 5.25, with a first quartile at 4.375 and a third quartile at 6.5, with a minimum of 3.5 and a maximum of 7. QA-based: The median is 6.5, with a first quartile at 6 and a third quartile at 6.625, with a minimum of 5 and a maximum of 7. Significant differences exist between No-AI and QA-based (p < 0.01), between Prompt-based and QA-based (p < 0.01), and between Prompt-based and QA-based (p < 0.05). H1-d: Difficulty for Task Initiation: No-AI: The median is 5, with a first quartile at 5 and a third quartile at 6, with a minimum of 1 and a maximum of 7. Prompt-based: The median is 3, with a first quartile at 1.75 and a third quartile at 4, with a minimum of 1 and a maximum of 5. QA-based: The median is 2, with a first quartile at 1 and a third quartile at 2, with a minimum of 1 and a maximum of 3. Significant differences exist between No-AI and Prompt-based (p < 0.01), between Prompt-based and QA-based (p < 0.01), and between No-AI and QA-based (p < 0.01). H1-e: Sense of Agency: No-AI: The median is 7, with a first quartile at 6.75 and a third quartile at 7, with a minimum of 6 and a maximum of 7. Prompt-based: The median is 4, with a first quartile at 4 and a third quartile at 4.25, with a minimum of 3 and a maximum of 5. QA-based: The median is 2.5, with a first quartile at 1.75 and a third quartile at 3.25, with a minimum of 1 and a maximum of 5. Significant differences exist between No-AI and Prompt-based (p < 0.01), between Prompt-based and QA-based (p < 0.01), and between No-AI and QA-based (p < 0.01). H1-e: Sense of Control: No-AI: The median is 7, with a first quartile at 7 and a third quartile at 7, with a minimum of 5 and a maximum of 7. Prompt-based: The median is 5, with a first quartile at 3.5 and a third quartile at 5, with a minimum of 1 and a maximum of 6. QA-based: The median is 3.5, with a first quartile at 2.75 and a third quartile at 4, with a minimum of 1 and a maximum of 6. Significant differences exist between No-AI and Prompt-based (p < 0.01), between Prompt-based and QA-based (p < 0.01), and between No-AI and QA-based (p < 0.01). H2: Perceived Quality of the Email by Evaluators: No-AI: The median is 5.22, with a first quartile at 4.43 and a third quartile at 5.42, with a minimum of 3.78 and a maximum of 5.83. Prompt-based: The median is 5.61, with a first quartile at 5.47 and a third quartile at 6.07, with a minimum of 4.56 and a maximum of 6.56. QA-based: The median is 5.81, with a first quartile at 5.43 and a third quartile at 5.99, with a minimum of 4.78 and a maximum of 6.50. Significant differences exist between No-AI and Prompt-based (p < 0.05) and between No-AI and QA-based (p < 0.01). H3-a: Perceived Impression of Participants by Evaluators: No-AI: The median is 4.17, with a first quartile at 3.31 and a third quartile at 4.63, with a minimum of 2.42 and a maximum of 5.67. Prompt-based: The median is 5.17, with a first quartile at 4.42 and a third quartile at 5.33, with a minimum of 3.92 and a maximum of 6.00. QA-based: The median is 4.92, with a first quartile at 4.60 and a third quartile at 5.42, with a minimum of 4.00 and a maximum of 5.58. Significant differences do not exist. H3-b: Psychological Distance: No-AI: The median overlap score is 5, with a first quartile at 2.75 and a third quartile at 6.25. The minimum score is 1, and the maximum score is 7. Prompt-based: The median overlap score is 4, with a first quartile at 2.75 and a third quartile at 4. The minimum score is 1, and the maximum score is 5. QA-based: The median overlap score is 1.75, with a first quartile at 3 and a third quartile at 3.5. The minimum score is 1, and the maximum score is 7. Significant differences exist between No-AI and QA-based (p < 0.05). Each box plot represents the spread of participant responses, with the whiskers showing the variability outside the upper and lower quartiles. The statistical differences (p-values) highlight significant findings between different experimental conditions.}
\end{figure*}
\section{Results of Study 1}
\red{Here, we first present the quantitative results of Study 1 for each research question. 
Subsequently, we include comments provided by the participants.}
\subsection{Participants' Email-Replying Process (RQ1)}
\label{sec:result1_RQ1}
\subsubsection{Efficiency of Replying to Emails (H1-a)}
\label{sec:result1_efficiency}
First, we compared the efficiency of replying to emails across three conditions.
After checking the data normality assumption with the Shapiro-Wilk test, the result of one-way repeated measures ANOVA showed that there was a significant difference in participants' efficiency of replying to emails across three conditions ($F[2, 22]=14.8$, $p<0.001$\red{, $\eta_p^2=0.57$}). 
Post-hoc analysis with Holm correction revealed that participants' efficiency of replying to emails in the QA-based condition was significantly higher compared to both the No-AI $(t(11), p=0.002\red{, d=1.38})$ and the Prompt-based $(t(11), p=0.046\red{, d=0.65})$ conditions.
Thus, H1-a was supported.
The QA-based approach enhanced participants’ email replying efficiency.

\subsubsection{Prompt Character Counts (H1-a)}
\label{sec:result1_prompt_character_counts}
In order to understand how participants wrote prompts differently, we calculated the prompt character counts.
After the Shapiro-Wilk test, the paired t-test revealed that participants in the QA-based condition typed significantly fewer characters in their prompts than those in the Prompt-based condition $(t(11), p=0.010\red{, d=0.90})$.

\subsubsection{Cognitive Load for Replying to Emails (H1-b)}
\label{sec:result1_cognitive_load}
The results of the Raw-TLX are shown in Fig.~\ref{fig_study1_efficiency_and_cognitiveLoad}.
According to the one-way repeated measures ANOVA with Greenhouse-Geisser correction, there was a significant difference in participants' cognitive load for replying to emails among the three conditions $(F[1.1, 12.1]=12.6, p=0.003\red{, \eta_p^2=0.53})$. 
Post-hoc analysis with Holm correction revealed that participants' cognitive load for replying to emails in the QA-based condition was significantly lower compared to both the No-AI $(t(11), p=0.008\red{, d=1.12})$ and Prompt-based $(t(11), p=0.018\red{, d=0.81})$ conditions.
Therefore, H1-b was supported.
The QA-based approach reduced participants’ cognitive workload while replying to the emails.

\subsubsection{\red{Difficulty in Understanding Email Content (H1-b)}}
\label{sec:result1_difficulty_in_understanding}
Additionally, H1-b was also supported by the questionnaire survey results (Fig.~\ref{fig_study1_questionnaire} H1-b).
The Friedman test revealed a significant difference among the three conditions in terms of understanding the sender's intent and requests $(\chi^2(2)=10.6, p=0.005\red{, W=0.44})$. 
Post-hoc analysis using the Durbin-Conover test with Holm correction showed that participants in the QA-based condition found it significantly easier to understand the sender's intent and requests compared to those in both No-AI $(p=0.003\red{, r=0.61})$ and Prompt-based $(p=0.005\red{, r=0.73})$ conditions.

\subsubsection{Satisfaction with Completing Task (H1-c)}
\label{sec:result1_satisfaction}
The results of the satisfaction with completing participants' tasks are shown in Fig.~\ref{fig_study1_questionnaire}, H1-c.
\red{The two items measuring satisfaction showed high internal consistency, with a Cronbach's Alpha of $0.889$.}
After checking the data normality assumption with the Shapiro-Wilk test, the result of one-way repeated measures ANOVA showed that there was a significant difference in participants' satisfaction with completing tasks across three conditions $(F[2, 22], p<0.001\red{, \eta_p^2=0.79})$. 
Post-hoc analysis with Holm correction revealed that participants' satisfaction with completing tasks in the QA-based condition was significantly higher compared to both the No-AI ($t(11)$, $p<0.001$\red{, $d=2.39$}) and the Prompt-based ($t(11)$, $p=0.029$\red{, $d=0.72$}) conditions.
Therefore, H1-c was supported.
The QA-based approach improved participants’ satisfaction with completing their tasks while replying to the emails.


\subsubsection{Difficulty in Initiating the Action for Replying to Emails (H1-d)}
\label{sec:result1_initiating}
The questionnaire survey results about participants' difficulty in initiating the action for replying to emails are shown in Fig.~\ref{fig_study1_questionnaire}, in H1-d.
According to the Friedman test, a significant difference in participants' difficulty in initiating the action for replying to emails was observed among the three conditions $(\chi^2(2)=19.8, p<0.001\red{, W=0.83})$.
Post-hoc analysis using the Durbin-Conover test with Holm correction revealed that participants in the QA-based condition perceived significantly higher barriers to initiating email response tasks than those in the No-AI $(p<0.001\red{, r=0.85})$ and Prompt-based $(p<0.001\red{, r=0.68})$ conditions.
Therefore, H1-d was supported.
The QA-based approach reduced participants’ difficulty in initiating the action to reply to emails.

\subsubsection{Sense of Agency and Control (H1-e)}
\label{sec:result1_agency}
The questionnaire survey results about a sense of agency and control are shown in Fig.~\ref{fig_study1_questionnaire}, H1-e.
The Friedman test revealed a significant difference among the three conditions for both the sense of agency $(\chi^2(2)=22.8, p<0.001\red{, W=0.95})$ and the sense of control $(\chi^2(2)=21.3, p<0.001$, $\red{W=0.89})$. 
Post-hoc analysis using the Durbin-Conover test with Holm correction showed that participants in the QA-based condition found that it significantly reduced their sense of agency compared to both the No-AI $(p<0.001\red{, r=0.88})$ and the Prompt-based $(p<0.001\red{, r=0.77})$ conditions.
Additionally, post-hoc analysis using the Durbin-Conover test with Holm correction showed that participants in the QA-based condition experienced a significantly reduction in their sense of control compared to both the No-AI $(p<0.001\red{, r=0.88})$ and the Prompt-based $(p=0.006\red{, r=0.56})$ conditions.
Thus, H1-e was supported.
The QA-based approach reduced participants’ sense of agency and sense of control while replying to the emails.

\subsection{Quality of the Email Responses (RQ2)}
\label{sec:result1_RQ2}
\subsubsection{Perceived Quality of the Email by Evaluators (H2)}
\label{sec:result1_quality}
In Fig.~\ref{fig_study1_questionnaire}, H2 shows the results regarding the quality of the emails.
\red{The Cronbach's Alpha of three items measuring the perceived quality of the email is $0.846$.}
After checking the data normality assumption with the Shapiro-Wilk test, the result of one-way repeated measures ANOVA showed that there was a significant difference in the perceived quality of the email across three conditions $(F[2, 22]=9.1, p=0.001\red{, \eta_p^2=0.45})$. 
Post-hoc analysis with Holm correction revealed that the perceived quality of the emails participants wrote in the QA-based condition was significantly higher compared to the No-AI $(t(11), p=0.005\red{, d=1.21})$ condition.
Thus, H2 was partially supported.
The QA-based approach improved the quality of the email responses compared to the No-AI condition.

\begin{table*}[t]
\caption{\red{Details of perceived quality of the emails. ($Mean\pm SD$)}}
\Description{The table presents the comparative evaluation of three methods, No-AI, Prompt-based, and QA-based, in terms of three key attributes: Politeness, Readability, and Meeting Demands. The results are displayed as mean scores with standard deviations. For Politeness, the No-AI method received a mean score of 4.39 ± 1.00, indicating lower politeness levels compared to the AI-based methods. The Prompt-based approach showed a significant improvement, scoring 5.65 ± 0.56, slightly outperforming the QA-based method, which scored 5.49 ± 0.51. In terms of Readability, the No-AI method achieved a score of 5.24 ± 0.79, again falling behind the AI-based methods. The Prompt-based approach scored 5.65 ± 0.69, while the QA-based method scored the highest at 5.78 ± 0.49, reflecting the most consistently readable outputs among the three. Finally, for Meeting Demands, the No-AI method scored 5.19 ± 0.76, which is comparatively lower than the AI-enhanced methods. The Prompt-based approach performed better with a score of 5.68 ± 0.60, but the QA-based method emerged as the best performer in this category, scoring 5.88 ± 0.60.}
\label{tab_study1_quality_of_emails}
\red{
\begin{tabular}{cccc}
\hline
             & Politeness                 & Readability                & Meeting Demands               \\ \hline
No-AI        & $4.39\pm1.00$ & $5.24\pm0.79$ & $5.19\pm0.76$ \\
Prompt-based & $5.65\pm0.56$ & $5.65\pm0.69$ & $5.68\pm0.60$ \\
QA-based     & $5.49\pm0.51$ & $5.78\pm0.49$ & $5.88\pm0.60$ \\ \hline
\end{tabular}
}
\end{table*}
\red{Tab.~\ref{tab_study1_quality_of_emails} shows the detailed results regarding the perceived quality of the emails across three evaluation dimensions (politeness, readability, and meeting demands).
These results further supported the partial acceptance of H2, showing that the AI-assisted approach tended to improve the email quality.} 


\subsection{Relationship between Participants and Their Counterpart (RQ3)}
\label{sec:result1_RQ3}
\subsubsection{Perceived Impression of Participants by Evaluators (H3-a)}
\label{sec:result1_impression}
The results of the perceived impression of the participants rated by another group of evaluators are shown in Fig.~\ref{fig_study1_questionnaire} H3-a.
\red{The two items assessing participants' impression as email senders showed high internal consistency, with a Cronbach's Alpha of $0.946$.}
After checking the data normality assumption with the Shapiro-Wilk test, the result of one-way repeated measures ANOVA showed that there was a significant difference in impression of the participants as an email sender across three conditions $(F[2, 22]=5.9, p=0.009\red{, \eta_p^2=0.35})$. 
Post-hoc analysis with Holm correction revealed that participants' impression in the QA-based condition was not significantly higher compared to both the No-AI $(t(11), p=0.058\red{, d=0.79})$ and the Prompt-based $(t(11), p=0.939\red{, d=0.02})$ conditions.
Thus, H3-a was not supported.
The QA-based approach didn't improve the impression of participants as email senders.

\subsubsection{Psychological Distance between Participants and Their Counterpart (H3-b)}
\label{sec:result1_psychological_distance}
The IOS result is shown in \red{Fig.~\ref{fig_study1_questionnaire}}.
The Friedman test showed a significant difference in psychological distance among the three conditions $(\chi^2(2)=7.47, p=0.024\red{, W=0.31})$.
Post-hoc analysis using the Durbin-Conover test with Holm correction revealed that IOS in the No-AI condition was significantly higher than in the QA-based condition $(p=0.021\red{, r=0.51})$. 
This result partially supports H3-b; however, because there was no significant difference between the Prompt-based and QA-based conditions \red{$(p=0.053, r=0.30)$}, H3-b was partially supported.

\subsection{\red{Qualitative Feedback}}
\label{sec:result1_interview}
\red{This section synthesizes qualitative feedback to provide further insights into participants' experiences across three conditions.
The interview comments were translated from Japanese into English.}
\subsubsection{\red{Participants' Email-Replying Process (RQ1)}}
\label{sec:result1_interview_RQ1}
\red{Feedback from participants confirmed that the QA-based condition functioned as expected, contributing to improvements in efficiency, a reduction in workload, and a lowering of barriers to task initiation compared to the other conditions.
\begin{enumerate}[]
    \item \textit{``In the QA-based condition, AI summarized key points through questions and highlighted relevant sections of the email body, which facilitated my understanding of the email and reduced my overall burden''} [P10].
    \item \textit{``In the QA-based condition, I could easily obtain the desired output even without the technical skills to create prompts''} [P6].
    \item \textit{``By saving the time needed to read the counterpart's text, the psychological barrier to starting the task was lowered''} [P5].
\end{enumerate}}

\red{On the other hand, we found that the QA-based condition led to a reduced sense of agency and control compared to the other conditions.
\begin{enumerate}[]
    \item \textit{``Since the AI prompted me with questions at the beginning, the mental effort required to start thinking about the task was eliminated, reducing the stress associated with initiating the work''} [P10].
    \item \textit{``By saving the time needed to read the counterpart's text, the psychological barrier to starting the task was lowered''} [P5].
\end{enumerate}}

\red{This aspect was also found to have the potential to negatively impact users' willingness to adopt the system in the future.
Participants noted that they preferred the QA-based condition \textit{``when time is limited or speed is important''} [P4] or \textit{when the email is of low importance''} [P5], but in other situations, they favored writing responses themselves.}

\subsubsection{\red{Quality of the Email Responses (RQ2)}}
\label{sec:result1_interview_RQ2}
\red{All participants stated that using AI improved their writing structure, politeness, and choice of words, ultimately enabling them to produce better overall responses. 
Furthermore, participants remarked, \textit{``Under the prompt-based condition, I might have overlooked the recipient's requests, but under the QA-based condition, I was able to craft responses with confidence''} [P2]. 
Additionally, one participant emphasized that \textit{``Under the QA-based condition, the AI even provided polite responses to matters where a reply was optional, such as acknowledging something with phrases like 'I Understood regarding XX, etc.'''} [P9], highlighting how the QA-based condition scaffolded user to construct a polite email in a formal setting.}

\subsubsection{Relationship between Participants and Their Counterpart (RQ3)}
\label{sec:result1_interview_RQ3}
\red{Participants shared differing views on how AI's involvement affected their psychological distance from their counterparts.
P2, P9, and P11 reported that the psychological distance they felt from the other person was directly related to the amount of effort they put in.
Furthermore, P6 noted that \textit{``especially under QA-based condition, I barely thought about the counterpart because I only selected options to create responses.''}
In contrast, P8 reported that \textit{``compared to composing replies myself, using AI allowed me to create messages that left a better impression on my counterpart, which made the relationship feel closer.''}
These results suggested that, on the one hand, AI's mediation can potentially increase the psychological distance between senders and receivers. 
On the other hand, it can also diminish the perceived distance from the sender due to effective impression management. Thus, we conducted a field study to further clarify the impact of AI on interpersonal relationships.}
\section{Method of Study 2}
Study 1 has revealed that the ResQ design effectively enhances user efficiency and reduces cognitive load during response tasks in formal settings.
\blue{To further examine how our QA-based system influenced users' actual email replying practice, we conducted a field experiment for Study 2.} 
\subsection{Field study}

In this five-day field experiment, we developed a prototype as a Chrome extension and asked participants to use the system to reply to emails using Gmail on a PC.
The extension detected the initiation of the reply task when participants clicked the ``Reply with AI'' button in the Gmail reply box. 
Upon confirming the task, the email content was sent to a remote server, a new reply editor opened, and a question with options appeared after a few seconds. 
Participants composed their replies, and upon pressing the Reply button, their text was directly reflected in the Gmail reply box. 
To ensure privacy, neither the email content nor the participants' responses were accessible to the experimenters or stored on the server.
\red{The specific implementation details and user interface are provided in the appendix.}


\subsection{Participants}
\begin{table*}[t]
\caption{\red{Backgrounds of participants in Study 2, including age, gender, job roles, frequency of AI tool usage, and use of AI for email purposes.}}
\Description{The table outlines the demographic information and AI usage patterns of nine participants in Study 2, including their age, gender, job roles, general AI tool usage, and the extent to which they utilize AI for email-related tasks. The participants include university students and office workers, with a mix of both male and female representatives, and their AI adoption varies from frequent to rare use. Participant P1 is a 28-year-old male office worker who uses AI tools daily, with 20–50\% of his email tasks supported by AI. P2, a 24-year-old female university student, frequently uses AI tools but does not employ them for email-related purposes. Similarly, P3, a 20-year-old female university student, frequently uses AI tools, with 20–50\% of her email tasks facilitated by AI. Participant P4, a 24-year-old female university student, engages in daily AI tool usage, relying on AI for 50–80\% of her email activities. P5, a 31-year-old male office worker, also uses AI tools daily, supporting 20–50\% of his email tasks. Likewise, P6, a 39-year-old female office worker, uses AI tools daily, with AI assisting in 20–50\% of her email-related tasks. In contrast, Participant P7, a 25-year-old male university student, rarely uses AI tools and does not employ them for email-related purposes. P8, a 23-year-old female office worker, frequently uses AI tools but applies them to less than 20\% of her email tasks. Lastly, P9, a 38-year-old male office worker, rarely engages with AI tools, with AI playing a role in less than 20\% of his email-related tasks.}
\label{tab_study2_participants_background}
\red{
\begin{tabular}{cccccc}
\hline
Participants & Age & Gender & Job       & AI Tool Usage&AI for Email Usage\\ \hline
P1           & 28  & M& Office Worker& Daily&20-50\%\\
P2           & 24  & F& Univ. Student& Frequently&Never\\
P3           & 20  & F& Univ. Student& Frequently&20-50\%\\
P4           & 24  & F& Univ. Student& Daily&50-80\%\\
P5           & 31  & M& Office Worker& Daily&20-50\%\\
P6           & 39  & F& Office Worker& Daily&20-50\%\\
P7           & 25  & M& Univ. Student& Rarely&Never\\
 P8           & 23  & F& Office Worker& Frequently&<20\%\\
P9& 38& M& Office Worker& Rarely&<20\%\\ \hline
\end{tabular}
}
\end{table*}
As shown in Tab.~\ref{tab_study2_participants_background}, nine participants (four males and five females, aged 20-39) were recruited via a local Japanese participant recruiting platform.
The average age of the participants was 28.0 (SD = 6.7)\blue{, and they reported engaging in more than three email communications per day on average.}
This study was approved by the ethical review board of the authors' institute.
The participants were paid approximately \$37 USD for participation.
The participants received a 30-minute explanation of the experiment and system, used the system for five days, and subsequently participated in a one-hour interview.

\subsection{Procedure}
Participants first read the study instructions and their right to participate, after which they consented to participate in the experiment. 
Next, they were provided with an explanation of the study's purpose and instructions on how to use the QA-based system. 
Following this, they installed the Chrome extension we developed and confirmed its functionality according to the provided instructions.
Participants were asked to use the system for five days, during which they were free to use it to reply to emails at any time. 
After the five-day period, a one-hour semi-structured interview was conducted. 
During the interview, participants were asked a series of questions, such as: \textit{``Can you tell us your overall impression of using the system?''} \textit{``How did your email replying practice change before and after using the system?''} \textit{``What changes did you notice in the emails you composed?''} and \textit{``How did your relationship with the communication counterpart change after using this system?''}
This study was conducted remotely with all participants.

\subsection{Data Analysis}
To analyze the interview data, we transcribed the interview recordings. 
We followed the thematic analysis method~\cite{braun_2006_thematicanalysis} to analyze the open-ended responses. 
One of the authors open-coded all relevant concepts that were related to our research questions, assigned labels that featured the concepts, and grouped labels into different themes. 
Next, the authors discussed the quotes and themes repeatedly.
Finally, the developed themes were compared and adjusted among all participants until they thoroughly covered the data.
As a result of the coding process, we identified four main themes: the email-replying process, the quality of the email responses, the relationship between the sender and recipient, and perceived risk.
\section{Results of Study 2}
\red{Tab.~\ref{tab_study2_participants_usage} presents the number of email replies composed using ResQ, along with the contexts in which it was used over five days.
We did not analyze usage frequency because participants reported avoiding using ResQ for emails for which they had privacy concerns.
Additionally, some participants refrained from using ResQ due to its availability only on PCs, as they frequently replied to emails via smartphone. 
Email frequency also varied among participants depending on their personal schedules (\textit{e.g.,} holidays).}

\red{Eight participants primarily used ResQ in formal workplace settings, while one (P9) used it only for informal exchanges. 
Because this was a field study, we could not limit participants to using ResQ only in formal contexts, though we instructed them to use it to reply to formal emails at the beginning. 
As a result, two participants (P3, P4) used ResQ to reply to both formal and informal emails, and P9 only used it for informal email exchanges. 
Hence, we excluded P9's data and only focused on analyzing the experience of P3 and P4 when they replied to formal emails using ResQ.}


\red{This section explains the results of interviews conducted with participants after they used the system, with the interview comments translated from Japanese into English.}
\begin{table*}[t]
\caption{Usage of the participants in Study 2. D1 through D5 represents the number of emails replied to using the system each day, from Day 1 to Day 5.}
\Description{The table illustrates the daily system usage of participants in Study 2, detailing the number of emails replied to each day (from Day 1 to Day 5) and the primary purposes for which the system was used. The participants represent a range of tasks, including work-related scheduling, academic communication, and informal interactions. Participant P1 primarily used the system for scheduling, task confirmations, and submissions related to work. His email activity was moderate, replying to 3 emails on Day 1, 1 email on Day 2, and 1 email on Day 5, with no emails replied to on Days 3 and 4. P2 focused on task management and communication related to research and university administration. She maintained consistent usage on Days 1 and 2, replying to 6 emails each day, did not use the system on Day 3, replied to 1 email on Day 4, and increased her activity to 5 emails on Day 5. Participant P3 engaged with the system for task management related to research with professors and informal contact with friends. Her usage was highest on Days 1 and 2, replying to 6 and 7 emails respectively. She replied to 2 emails on both Days 3 and 4 and did not use the system on Day 5. P4 used the system for scheduling related to club activities, informal contact with friends, and inquiries with a museum abroad. She consistently replied to 6 emails on both Days 1 and 2, 5 emails on Day 3, 4 emails on Day 4, and 6 emails on Day 5. Participant P5’s primary usage involved meeting planning and confirmations related to work. He replied to 2 emails on Day 1, 3 emails on Day 2, 5 emails on Day 3, and 1 email on both Days 4 and 5. P6 utilized the system for scheduling and confirmations related to work, friends, and event organizers. Her activity peaked on Days 2 and 3, replying to 6 emails each day, followed by 3 emails on both Days 1 and 4 and 5 emails on Day 5. Participant P7 focused on scheduling and progress management related to research and business trips. His usage was irregular, with 4 emails replied to on Day 1, none on Day 2, 6 emails on Day 3, none on Day 4, and 2 emails on Day 5. Lastly, P8 primarily used the system for progress management and administrative confirmations related to work. Her activity started with 5 emails on Day 1, decreasing to 2 emails on Day 2, 1 email on Day 3, and 5 emails on Day 4, with no usage recorded on Day 5.}
\label{tab_study2_participants_usage}
\begin{tabular}{ccccccl}
\hline
\multirow{2}{*}{} & \multicolumn{5}{c}{Daily System Usage} & \multicolumn{1}{c}{\multirow{2}{*}{Main Usage}}                                                      \\ \cline{2-6}
                  & D1     & D2    & D3    & D4    & D5    & \multicolumn{1}{c}{}                                                                                 \\ \hline
P1                & 3      & 1     & 0     & 0     & 1     & Scheduling, task confirmations, and submissions related to work                                      \\
P2                & 6      & 6     & 0     & 1     & 5     & Task management and communication related to research and university administration                  \\
P3                & 6      & 7     & 2     & 2     & 0     & Task management related to research with professors, informal contact with friends                   \\
P4                & 6      & 6     & 5     & 4     & 6     & Scheduling related to club activities, informal contact with friends, inquiries with a museum abroad \\
P5                & 2      & 3     & 5     & 1     & 1     & Meeting planning and confirmations related to work                                                   \\
P6                & 3      & 6     & 6     & 3     & 5     & Scheduling and confirmations related to work, friends, and event organizers                          \\
P7                & 4      & 0     & 6     & 0     & 2     & Scheduling and progress management related to research and business trips                            \\
P8                & 5      & 2     & 1     & 5     & 0     & Progress management and administrative confirmations related to work                                 \\ \hline
\end{tabular}
\end{table*}
\subsection{Participants' Email-Replying Process (RQ1)}
\subsubsection{Improved Perception of Efficiency and Workload}
\label{sec:result2_efficiency}
Participants reported that their perception of workload and work efficiency improved due to the support from ResQ.
Specifically, participants noted that ResQ's support helped clarify the topics they needed to address in the email. 
Participants explained that \textit{``Normally, when writing, I need to process multiple tasks simultaneously to ensure my intentions are appropriately expressed. However, [With ResQ,] replying to emails was divided into two different sub-tasks, answering questions and polishing emails with diverse expressions. As a result, I felt that the cognitive load was reduced.''} [P7], and \textit{``it felt like creating an email was as simple as answering a survey''} [P6].
Additionally, particularly when the counterparts' message was long, participants reported that the listing of requests as questions allowed them to \textit{``easily understand the content of the email''} [P3], with another participant noting that \textit{``I can quickly make decisions on what to reply [with ResQ]''} [P6]. 
Furthermore, compared to other AI tools like ChatGPT, the QA-based approach enabled participants to communicate their intentions more efficiently without extensive typing.
As one participant described,\textit{``[Writing with ResQ] made it easier to reflect my intentions while replying to the email''} [P4], while another participant added that \textit{``I could create the expected reply without even having to type on the keyboard''} [P6].
\red{Additionally, all participants expressed increased satisfaction with the quality of the responses they wrote (for more details, see Sec.~\ref{sec:result2_quality}) and responded positively to the question, \textit{``What is your overall impression of using ResQ?''} and expressed a desire to continue using the system in the future.}
Participants also mentioned that being able to craft clearer messages more quickly than before resulted in \textit{``greater confidence in the reply process and a more positive perception of the task''} [P8]. 
Additionally, a different participant expressed, \textit{``I felt joy in meeting societal expectations competently''} [P2].
These increases in achievement and confidence led participants to report that their \textit{``perception of the reply task became more positive''} [P8], and they felt \textit{``more motivated to engage actively in email responses''} [P3]. 

\subsubsection{Reduced Difficulty in Initiating the Action for Replying to Emails}
\label{sec:result2_initiating_the_action}
Participants reported that ResQ's support lowered the barrier to starting tasks, reducing procrastination in replying to emails.
One participant shared that they previously \textit{``felt reluctant to engage in replying due to the burden of the task''}, but with ResQ, \textit{``I felt motivated because I can complete the task quickly''} [P3]. 
Another participant noted that \textit{``I became able to craft replies to any email easily, so I could respond even on days when I was tired or when I would typically postpone replying to long emails''} [P6].
Participants also reported that using AI to initiate the task motivated them to start replying to emails without procrastinating. 
One participant explained that \textit{``just pressing a button prompts the AI to ask questions''}[P4], which led them to \textit{``delegate the initial steps entirely to the system''} [P3]. 
This reduction in the burden of the initial stage was cited as a key factor in lowering the barrier to starting to reply to emails. 

\subsubsection{Reduced Sense of Agency and Control}
\label{sec:result2_agency_control}
\red{Three out of eight participants (P3, P5, and P6) reported a decreased sense of agency and control while replying to emails with ResQ.}
They attributed this to several factors: one participant mentioned that their perception shifted \textit{``from that of an author to that of an editor''} [P5], which reduced the workload of replying but made the process feel \textit{``like an assembly line''} [P3], while another expressed, \textit{``I ended up using words or expressions I normally wouldn't [use in the email]''} [P6].
In contrast, for those who reported no change in their sense of agency or control (five participants), they explained that this was because  \textit{``the email content was strongly related to me''} [P7], and they \textit{``checked the content carefully''} [P7] or \textit{``modified words that I wouldn't normally use to the ones I would use''} [P2], leading them to feel that their \textit{``active involvement [to reply to the email] was indispensable''} [P8]

\subsection{Increased Perceived Quality of the Email (RQ2)}
\label{sec:result2_quality}
Participants reported that they felt the quality of their emails had improved. 
Participants explained that, in the process of creating responses, they were most concerned with \textit{``politeness in language, such as expressions and greetings''} [P6], and mentioned that ResQ provides support in these areas.
Participants reported that \textit{``it was helpful to have phrases that would have taken time to come up with on their own, expressions of apology and gratitude, and additional words of consideration for the other person''} [P2], \textit{``there was no need to think about the opening and closing greetings''} [P8], and \textit{``there were no typos or omissions at all''} [P5].
Additionally, participants mentioned that ResQ helped reduce the likelihood of overlooking requests in the emails they received. 
One participant shared, \textit{``Previously, when a single email contained multiple requests, I sometimes missed responding to all of them, but the questions provided by ResQ helped improve this''} [P6].
Participants attributed this improvement to the fact that ResQ \textit{``secured time to focus on understanding the recipient's requests and responding to them''} [P1], and the questions generated by ResQ \textit{``helped me ensure that nothing was overlooked in the content''} [P7]. 
Furthermore, participants reported that responding to AI-generated questions encouraged them to include details they would normally omit, resulting in more polite and comprehensive responses. 
One participant described an email regarding event attendance and multiple confirmations, explaining that while they would usually reply with something like \textit{``I will attend, thank you''}, answering the AI's questions led to a response where \textit{``each of the recipient's requirements was addressed more carefully''} [P2], ultimately leading to a more courteous email.

\subsection{Relationship between Participants and Their Counterpart (RQ3)}
\subsubsection{Enabling a Positive Self-Presentation as an Email Sender}
\label{sec:result2_self-presentation}
The participants reported feeling they could make a good impression on others using ResQ. 
They attributed this to improvements in the quality of their writing, shorter response times, and increased frequency of replies. 
One participant mentioned, \textit{``I could answer the other person's questions clearly, and the writing became more polished, making it easier for them to read''} [P5]. 
The participant also mentioned that \textit{``I felt the individuality of the email reply had faded''} but added that \textit{``I never intended to express individuality in my emails to begin with, so even if it was lost, it wasn't an issue as long as it felt natural to the recipient''} [P5].
Another participant shared that when they met a professor with whom they had communicated via ResQ, the person remarked, \textit{``Your emails have become more polished.''} 
They further elaborated, \textit{``I was particularly complimented on how much more understandable the structure of my emails has become''} [P3].
Additionally, this participant noted, \textit{``Previously, I would often respond to long emails with just, 'I'll get back to you later,' because reading through and thinking about a proper reply was tedious. However, [with ResQ's support,] I've started responding immediately instead of postponing. As a result, I've been assigned more tasks than before.''}

\subsubsection{Psychological Distance between Participants and Their Counterpart}
\label{sec:result2_psychological_distance}
Participants had mixed opinions regarding the psychological distance they perceived from their counterparts.
Those who felt the decreased psychological distance between themselves and their counterparts attributed this to the positive impression they believed they made on their counterparts. 
One participant reported that sending well-crafted emails quickly led to \textit{``a stronger sense of reassurance in [formal] communication''} [P5], while another participant noted that \textit{``[When I asked the museum staff a question,] I noticed that when I replied immediately after receiving a message from the other person, they responded quickly in return. When we communicated with such a good rhythm, I felt a strong sense of closeness towards the counterpart''} [P4].
In contrast, participants who felt the increased psychological distance mentioned a strong awareness that their replies were mediated by a system and the use of words they would not usually choose. 
One participant gave an example of communication with their university professor, stating, \textit{``While I know the counterpart typed their emails manually, I felt that using AI made the conversation more superficial, which weakened our relationship''} [P3]. 
Participants also shared that they tended to forget about the email exchange with their counterparts due to the increased psychological distance.
One participant mentioned, \textit{``I found the email content easy to understand while working on it [with ResQ], but I felt it was difficult to retain our email exchange in long-term memory. When that counterpart [who is my professor] asked me, 'What happened with that issue? [that had been mentioned in our email]' there were times I couldn’t remember, which made me feel anxious''} [P3].

\subsection{Perceived Risks}
\label{sec:result2_risks}
Participants expressed concerns about the potential risks that ResQ might pose in the future. 
They expressed concerns about potential declines in their abilities and the risk of becoming overly dependent on AI, which could lead to carelessness in responding to work-related emails.
One participant explained, \textit{``I worry that the skills I've developed from composing emails myself might deteriorate''} [P8]. 
Another participant voiced concerns that \textit{``the advancement and usage of AI [in this context] might erode our ability to overcome psychological barriers''} [P2], fearing a decline in their interpersonal communication skills.
Additionally, participants raised the issue of over-reliance on AI, with one participant noting, \textit{``Given my trust in AI, I might eventually stop reviewing the content of the emails I send or the emails I receive''} [P8]. 
This reflects their concern about the potential for becoming overly dependent on AI-generated text in the future.

\section{Discussion}
\red{Through a controlled experiment (Study 1) and a field study (Study 2), we investigated the impact of the LLM-powered QA-based approach on both senders and receivers.
In this section, we discuss the findings (Fig.~\ref{tab_summary}) of the research and the key considerations for designing QA-based systems.}
\begin{table*}[t]
\caption{Research Questions and Key Findings}
\label{tab_summary}
\centering
\begin{tabular}{>{\raggedright\arraybackslash}p{0.08\linewidth}>{\raggedright\arraybackslash}p{0.28\linewidth}>{\raggedright\arraybackslash}p{0.28\linewidth}>{\raggedright\arraybackslash}p{0.28\linewidth}}
\hline
 & \textbf{RQ1: How does a QA-based response-writing support approach affect workers’ email-replying process?} & \textbf{RQ2: How does a QA-based response-writing support approach affect the quality of the email response?} & \textbf{RQ3: How does a QA-based response-writing support approach affect the perceived relationship between email sender and recipient?} \\ \hline
\textbf{Key Findings} & 1. QA-based approach \textbf{reduced workload} for email comprehension and prompt creation and \textbf{improved work efficiency}. (H1-a, supported; H1-b, supported, Sec.~\ref{sec:result1_efficiency},~\ref{sec:result1_prompt_character_counts},~\ref{sec:result1_cognitive_load},~\ref{sec:result1_difficulty_in_understanding},~\ref{sec:result1_interview_RQ1},~\ref{sec:result2_efficiency}) 

2. QA-based approach \textbf{reduced the difficulty} of initiating the email replying task. (H1-d, supported, Sec.~\ref{sec:result1_initiating},~\ref{sec:result1_interview_RQ1},~\ref{sec:result2_initiating_the_action})

3. QA-based approach \textbf{decreased the sense of agency and control}. (H1-e, supported, Sec.~\ref{sec:result1_agency},~\ref{sec:result1_interview_RQ1},~\ref{sec:result2_agency_control})

4. QA-based approach \textbf{improved satisfaction} with the emails they wrote and willingness to use ResQ in the future. (H1-c, supported, Sec.~\ref{sec:result1_satisfaction},~\ref{sec:result1_interview_RQ1},~\ref{sec:result2_efficiency})& Writing emails with QA-based approach and Prompt-based approach led to \textbf{increased email quality} than No-AI condition. (H2, partially supported, Sec.~\ref{sec:result1_quality},~\ref{sec:result1_interview_RQ2}~\ref{sec:result2_quality})& 1. Writing emails with QA-based approach \textbf{\blue{did not lead to improved perceived impression of users by their counterparts}}. (H3-a, not supported, Sec.~\ref{sec:result1_impression},~\ref{sec:result2_self-presentation})

2. Writing emails with QA-based approach led to \textbf{increased psychological distance} between users and their counterparts than No-AI condition. (H3-b, partially supported, Sec.~\ref{sec:result1_psychological_distance},~\ref{sec:result1_interview_RQ3},~\ref{sec:result2_psychological_distance})\\ \hline
\end{tabular}
\Description{This table summarizes the three research questions (RQs) investigated in the study and highlights the key findings associated with each. RQ1: How does a QA-based response-writing support approach affect workers’ email-replying process? Key Findings: 1. The QA-based approach reduced workload for email comprehension and prompt creation, leading to improved work efficiency. This supports hypotheses H1-a and H1-b. 2. It reduced the difficulty of initiating the email replying task, supporting hypothesis H1-d. 3. The approach decreased the sense of agency and control among users, supporting hypothesis H1-e. 4. Users experienced improved satisfaction with the emails they wrote and showed a greater willingness to use ResQ in the future, supporting hypothesis H1-c. RQ2: How does a QA-based response-writing support approach affect the quality of the email response? Key Findings: Writing emails using both the QA-based and prompt-based approaches led to an increase in email quality compared to the No-AI condition. This partially supports hypothesis H2. RQ3: How does a QA-based response-writing support approach affect the perceived relationship between email sender and recipient? Key Findings: 1. Writing emails with QA-based approach did not lead to improved perceived impression of users by their counterparts, meaning hypothesis H3-a was not supported. 2. Writing emails with the QA-based approach led to an increase in psychological distance between users and their counterparts compared to the No-AI condition. This partially supports hypothesis H3-b.}
\end{table*}

\subsection{Impact of the QA-based Approach}
\subsubsection{Enhancing Efficiency and Reducing Cognitive Load}
Our studies indicate that the QA-based approach improves efficiency and \blue{suggests a reduction in} cognitive load when composing email replies (Sec.~\ref{sec:result1_efficiency},~\ref{sec:result1_prompt_character_counts}, \ref{sec:result1_cognitive_load},~\ref{sec:result1_difficulty_in_understanding},~\ref{sec:result1_interview_RQ1},~\ref{sec:result2_efficiency}). 
One possible explanation is that the QA-based approach helps users focus on the most relevant details, simplifying email comprehension compared to prompt-based methods.
Additionally, the QA-based approach reduces the burden of prompt creation by partially replacing the task of crafting prompts with the simpler task of answering questions.
\blue{Our finding suggests that future email systems could use this QA-based approach to mediate the email exchange process.}

\subsubsection{Potential Reduction in Sense of Agency and Control}
\red{While the QA-based approach enhanced users' efficiency, our studies also revealed a potential trade-off in users' sense of agency and control (Sec.~\ref{sec:result1_agency},~\ref{sec:result1_interview_RQ1},~\ref{sec:result2_agency_control}).
Some participants reported a decreased sense of authorship, feeling more like editors than creators of their emails. 
This reduction in agency may be due to the diminished amount of text input required from the user, as the AI takes a more active role in content generation.
Moreover, we found that the sense of agency influenced users' preferences for future usage.
Among those participants who still maintained their sense of agency, we found that they tended to actively review and modify the AI-generated content to reflect their personal style and intentions.
\blue{This suggests that even when AI intervention is substantial, users can maintain a sense of authorship by actively engaging with and refining the AI's suggestions.}
}
\blue{To optimize users' level of agency, adapting the degree of AI intervention in the email construction process can be helpful.
For instance, by adjusting the number and type of AI-generated questions or varying the levels of AI-generated suggestions~\cite{Fu2023Comparing}, ranging from word-level to message-level.} 


\subsubsection{Possibility of Improving Relationship between Email Sender and Recipient}
\red{Our studies yielded mixed results regarding the impact of the QA-based approach on the psychological distance between users and their counterparts (Sec.~\ref{sec:result1_psychological_distance},~\ref{sec:result1_interview_RQ3},~\ref{sec:result2_psychological_distance}). 
Some participants reported that they were able to send emails more quickly and with high quality, which in turn led to faster responses from others and a reduced sense of distance in their interactions.
In contrast, other participants experienced an increased sense of distance, which has also been reported in the previous studies~\cite{Fu2023Comparing,arnold2020predictive}. 
They noted that the reduced communication effort and the use of unfamiliar language made interactions feel less personal or authentic.
The degree of the perceived distance may depend on factors such as the nature of the relationship (\textit{e.g.,} colleagues vs. friends), the user's reliance on AI-generated language, and individual preferences regarding AI-mediation in communication.}

\subsection{Opportunities and Challenges of Introducing QA-Based Approach}
Our results indicate that the QA-based approach 
is particularly useful in situations where speed and high-quality responses are prioritized over email personality or a strong sense of personal agency. 
Contexts such as business, customer service, and technical support can greatly benefit from the QA-based approach, as they often require efficient and structured communication.

However, for more delicate or personal email exchanges, users may prefer more tailored interventions. 
In such situations, users can adjust the level of involvement of AI intervention.
Furthermore, there is a risk that users could become overly reliant on technology to mediate their interpersonal communication. 
Our interviews revealed that users might become accustomed to trusting AI-generated questions and drafts due to the efficient outcomes. 
Consequently, they may become less diligent in reading the emails they receive or in reviewing the responses they send carefully.
This over-reliance could lead to miscommunication or the omission of important details, thus undermining the primary goal of using AI to improve communication efficiency.
Future research should explore how different levels of AI-mediated intervention can be designed to influence users’ sense of agency and email construction behavior for various communication purposes.

\subsection{Limitations and Future Work}
While it is evident that the QA-based approach positively impacted users' workload, the quality of the emails they produced, and their relationship with recipients in formal email responses, this study had several limitations.
Though we tried to use a mixed-method study to triangulate the findings from the control experiment and field study, we acknowledged that the quantitative results could be limited.
Because of privacy concerns, we were unable to access participants' email content, and as a result, we could not gather users' behavioral data. 
This includes information such as how they edited the prompts, the amount of time they dedicated to responding to emails, or how ResQ influenced the language they used in their actual email communications.
We encourage researchers to explore alternative research methods for capturing users' behavioral data in email exchanges in the wild to enrich the understanding of QA-based approaches in AI-mediated communication.

\red{Second, the effectiveness of the QA-based approach may vary depending on the specific characteristics of the emails. 
We conducted Study 1 using emails on a variety of topics within formal scenarios to examine the impact of the QA-based approach. 
However, its effectiveness may differ based on characteristics such as the formality of the situation, the politeness of the email, its importance, or the relationship between the sender and recipient.
Therefore, future research could explore how these specific email characteristics influence the effectiveness of QA-based approaches, potentially tailoring AI-mediated tools to different communication contexts.}

\blue{Third, the study was conducted with participants from a single cultural background, which could limit the generalizability of our findings. 
Although we contributed to a new understanding for populations from non-Western countries~\cite{WEIRD_CHI21}, we acknowledge that the practice of email exchange differs across cultures~\cite{Robertson2021ICant}.
Further studies are encouraged to examine whether similar results would be obtained among users from diverse cultural backgrounds or in cross-cultural email exchanges.}



\red{Fourth, while this study focused on a QA-based approach driven by LLMs, future research could explore alternative methods of question generation to deepen our understanding of QA-based AI assistance. 
For instance, comparing the LLM-powered system with approaches utilizing rule-based question generation or manually prepared questions and options may help disentangle the effects of algorithmic sophistication from the inherent benefits of structuring communication as QA. 
This may potentially clarify whether the AI placebo or nocebo effect~\cite{kloft2024aiplacebo} exists in AI-mediated communication.
Examining these different methods could offer further insights into when and why the QA-based approach excels and guide the design of more tailored systems that accommodate a wide range of communication tasks and user needs.}

\red{Fifth, while this study demonstrated the effectiveness of the QA-based approach with initial design considerations (Sec.~\ref{sec:Proposed_Approach}), future research could explore tailoring these questions to specific communication goals or contexts. 
For example, designers or instructors could adjust factors such as the number of questions, their difficulty level, or their thematic focus to improve the user's understanding of challenging content. 
By iterating on the design to explore how different dimensions of question can affect communication outcomes, future work can better guide the QA-based approach.}

\section{Conclusion}
\red{In formal email communication, users are often required to read detailed (lengthy or complex) emails. 
Crafting appropriate responses to such emails is time-consuming and may lead to overlooked sender requests or delayed responses, causing communication issues.}
Thus, we propose \red{QA-based approach}, which leverages LLM-based question generation to help users create efficient and high-quality replies by generating multiple question-answer pairs related to the received email content.
\red{To examine the comprehensive impact of the QA-based approach on both email senders and recipients, we conducted controlled and field experiments using our prototype system, \textit{ResQ}.
Our findings demonstrate that structuring email content into question-answer pairs improves efficiency, reduces cognitive load, and lowers barriers to initiating responses. 
Additionally, this approach enhances email quality and may leave a better impression on recipients.
However, our findings also revealed challenges, including a potential reduction in user agency and an increased psychological distance in communication. 
These trade-offs emphasize the need for adaptive designs that balance efficiency with personalization and user control.
Future research should investigate the long-term effects of such systems on user behavior, cross-cultural differences in adoption, and the effectiveness of the QA-based approach across varying email characteristics}

\begin{acks}
This work was supported by JSPS KAKENHI (JP24H00742 and JP24H00748).
We thank all the participants for their interest and involvement in this study.
We also appreciate the reviewers for their constructive feedback, which helped us refine this work.
\end{acks}


\appendix

\lstset{
  backgroundcolor=\color{gray!20}, 
  basicstyle=\ttfamily\footnotesize, 
  breaklines=true,                 
  frame=single,                    
  framerule=0pt,                   
  xleftmargin=5pt, xrightmargin=5pt 
}

\section{Prompt}
In this section, we list the full prompts given to the LLM, which were used in this paper.
\subsection{Prompt to Generate Questions}
\begin{lstlisting}
###Instruction###
You are assisting the audience who has received an email and needs to respond.
You're like a secretary for your audience, asking them questions and creating email responses on their behalf based on their answers.
Your goal is to make it as clear as possible what and how your audience wants to answer in response to all requirements of the email.
Therefore, you must create as specific questions as possible.
Specifically, you will assist your audience in composing emails by following 3 steps:

Step 1: Create Questions:  
    To achieve your goal, you must create well-thought-out questions without omission by considering the sender's intent and requirements. 
    The number and content of the questions must be determined with this in mind.
Step 2: Receive Answers:  
    Ask your audience the questions you created and collect their responses. 
    These will guide the crafting of the reply.
Step 3: Propose a Reply:  
    Based on the answers received, suggest a reply that your audience can edit and send.
    From now on, you will perform step 1.  

You must consider the following 7 matters in generating your response.

1. You must create questions with choices for your audience and output the results in JSON format.
2. The questions must be created in the native language of your audience.
3. If necessary, your audience can write any free answers to your questions, so you will be penalized if you create an "other" option.
4. In 'corresponding_part', you must quote a part of the provided 'Incoming Mail' verbatim. That is, output corresponding_part = IncomingMail_HTML[x:x+h]. 
5. You must quote spaces, `<br>`, periods, and commas exactly as in the provided 'Incoming Mail'. 
6. You will be penalized if you edit or combine multiple parts of the 'Incoming Mail' for your questions.
7. You will be penalized if you create unhelpful questions to compose a reply. You must keep the number of questions to a minimum.

I'm going to tip $100 for a better solution!
Ensure that your output is unbiased and avoids relying on stereotypes.

###Output JSON Format###
{
    "questions": [
        {
            "id": "1",
            "question": "Will you participate in the event on October 24th?", 
            "choices": ["Yes", "No"], 
            "corresponding_part": "We will hold an event on October 24th."
        },
        {
            "id": "2",
            "question": "Please select the available dates (multiple selections possible).", 
            "choices": ["July 10th", "July 11th", "July 12th", "July 13th", "July 14th", "July 15th", "July 16th"], 
            "corresponding_part": "Please let us know your available dates within a week."
        }
    ]
}
\end{lstlisting}

\subsection{Prompt to Generate Reply Draft}
Below is the prompt used to generate a balanced-length description.
\begin{lstlisting}
Please provide a draft reply to the sender of this email on behalf of the user.
\end{lstlisting}

\section{\blue{Order Effect in Study 1}}
\blue{We conducted analyses to examine whether the order of conditions influenced various dependent variables. 
Table~\ref{tab_ordereffects} summarizes the results of the order effect and its interaction with the condition. 
Depending on the nature of the data, we employed either a Mixed-Design ANOVA or the Aligned Rank Transform (ART) method.}

\begin{table*}[t]
\caption{Order effects in Study 1 (* indicates significance at the 0.05 level).}
\Description{The table presents the order effects observed in Study 1, examining whether the order of conditions influenced various measurements. It reports p-values for the main effect of Order and the interaction effect between Condition and Order, with asterisks indicating statistical significance at the 0.05 level. The statistical method used for each measurement is also specified, including Mixed-Design ANOVA and Mixed-Design ANOVA with Aligned Rank Transform (ART). For Efficiency of Replying to Emails, the Order effect has a p-value of 0.643, and the Condition × Order interaction has a p-value of 0.454, analyzed using Mixed-Design ANOVA. Prompt Character Count shows an Order effect p-value of 0.052, close to significance, while the Condition × Order interaction has a p-value of 0.186, also analyzed using Mixed-Design ANOVA. Raw TLX, which measures subjective workload, has a significant Order effect (p < 0.05), indicating that workload perception is influenced by order, whereas its Condition × Order interaction is non-significant (p = 0.978), analyzed using Mixed-Design ANOVA. For Difficulty in Understanding Email Content, the Order effect p-value is 0.188, and the Condition × Order interaction p-value is 0.232, analyzed using Mixed-Design ANOVA with ART. Satisfaction with Completing Tasks shows non-significant effects, with an Order effect p-value of 0.348 and a Condition × Order interaction p-value of 0.175, analyzed using Mixed-Design ANOVA. Similarly, Difficulty for Task Initiation has an Order effect p-value of 0.131 and a Condition × Order interaction p-value of 0.515, analyzed using Mixed-Design ANOVA with ART. Psychological perceptions are also analyzed. Sense of Agency shows non-significant results, with an Order effect p-value of 0.982 and a Condition × Order interaction p-value of 0.911, analyzed using Mixed-Design ANOVA with ART. Sense of Control similarly has an Order effect p-value of 0.825 and a Condition × Order interaction p-value of 0.871, analyzed using Mixed-Design ANOVA with ART. Regarding Perceived Quality of the Email, the Order effect is 0.930, and the Condition × Order interaction is 0.433, analyzed using Mixed-Design ANOVA. Perceived Impression of Participants also shows non-significant effects, with an Order effect p-value of 0.963 and a Condition × Order interaction p-value of 0.481, analyzed using Mixed-Design ANOVA. Finally, Psychological Distance presents an Order effect p-value of 1.000, indicating no effect of order, but its Condition × Order interaction is significant (p < 0.05), suggesting that condition effects on psychological distance are influenced by order. This measurement is analyzed using Mixed-Design ANOVA with ART.}
\label{tab_ordereffects}
\blue{
\begin{tabular}{cccc}
\hline
Measurements                              & Order (p-value) & Condition $\times$ Order (p-value) & Statistical Method          \\ \hline
Efficiency of Replying to Emails          & 0.643            & 0.454                        & Mixed-Design ANOVA          \\
Prompt Character Count                    & 0.052           & 0.186                       & Mixed-Design ANOVA          \\
Raw TLX                                   & $<0.05$*         & 0.978                       & Mixed-Design ANOVA          \\
Difficulty in Understanding Email Content & 0.188           & 0.232                       & Mixed-Design ANOVA with ART \\
Satisfaction with Completing Tasks        & 0.348           & 0.175                       & Mixed-Design ANOVA          \\
Difficulty for Task Initiation            & 0.131           & 0.515                       & Mixed-Design ANOVA with ART \\
Sense of Agency                           & 0.982           & 0.911                       & Mixed-Design ANOVA with ART \\
Sense of Control                          & 0.825           & 0.871                       & Mixed-Design ANOVA with ART \\
Perceived Quality of the Email            & 0.93            & 0.433                       & Mixed-Design ANOVA          \\
Perceived Impression of Participants      & 0.963           & 0.481                       & Mixed-Design ANOVA          \\
Psychological Distance                    & 1.000           & $<0.05$*                     & Mixed-Design ANOVA with ART \\ \hline
\end{tabular}
}
\end{table*}

\blue{The results indicate that the order effect was not significant for most dependent variables. 
However, a significant effect was observed for Raw TLX ($p = 0.041$), suggesting that task load perception may have been influenced by the presentation order. 
Additionally, a significant interaction effect between condition and order was found for IOS ($p = 0.043$), indicating that the order of presentation might have impacted this specific measure.}

\section{Future Preference in Study 1}
\begin{figure*}[ht]
\centering
\includegraphics[width=\textwidth]{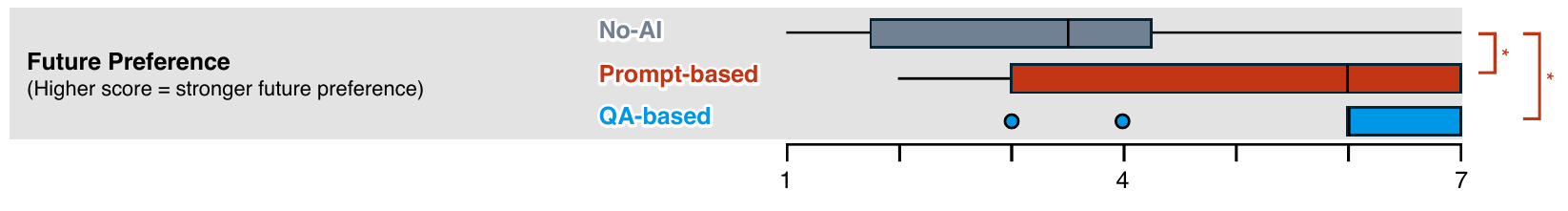}
\caption{Participants' future preferences in Study 1. Significant differences between conditions were identified through post-hoc analysis following the Friedman test (* indicates significance at the 0.05 level).}
\label{fig_study1_preference}
\Description{The figure displays participants' future preferences using box plots for three experimental conditions: No-AI, Prompt-based, and QA-based. For the No-AI condition, the median overlap score is 3.5, with the first quartile at 1.75 and the third quartile at 4.25. The scores range from a minimum of 1 to a maximum of 7. For the Prompt-based condition, the median overlap score is 6, with the first quartile at 3 and the third quartile at 7. The scores range from 2 to 7. For the QA-based condition, the median overlap score is 6, with the first quartile at 6 and the third quartile at 7. The scores range from 3 to 7. Significant differences were observed between the No-AI condition and both the Prompt-based and QA-based conditions (p < 0.05).}
\end{figure*}
\red{We evaluated participants' preferences for future use across all conditions using a 7-point Likert scale.
Participants rated their agreement with the statement, ``I would prefer to use this approach for replying to emails in the future,'' where 1 indicates strongly disagree, 4 indicates neutral, and 7 indicates strongly agree.}

\red{The questionnaire survey results about participants' future preferences are shown in Fig.~\ref{fig_study1_preference}.
According to the Friedman test, a significant difference in participants' future preferences was observed among the three conditions $(\chi^2(2)=8.8, p=0.012, W=0.37)$.
Post-hoc analysis using the Durbin-Conover test with Holm correction revealed that participants would prefer responding in the QA-based condition compared to the No-AI condition $(p=0.012, r=0.67)$.
However, no significant difference was found between the Prompt-based and QA-based conditions $(p=0.800, r=0.27)$.}

\section{Technical Details of ResQ}
\begin{figure*}[ht]
\centering
\includegraphics[width=\textwidth]{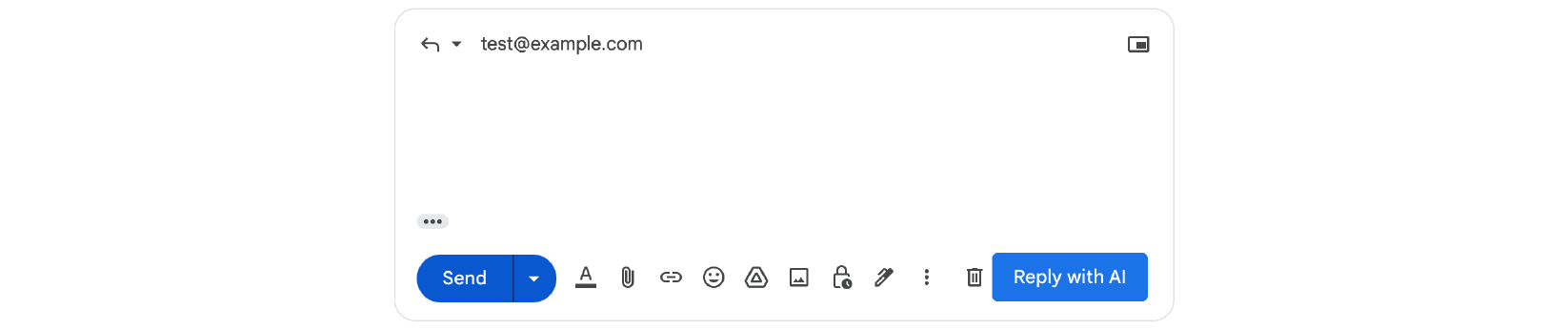}
\caption{UI of the Gmail Reply Box with the ``Reply with AI’’ Feature, used in Study 2. Pressing the ``Reply with AI’’ button opens the window shown in Fig.~\ref{fig_interface}}
\label{fig_UI}
\Description{This figure illustrates the user interface of the Gmail reply box as enhanced by the prototype system. The Reply with AI button, shown in blue on the right-hand side of the toolbar, allows users to activate the AI-assisted reply generation feature. When the button is clicked, the system extracts the email content and opens the window shown in Fig.~\ref{fig_interface}. The standard Gmail toolbar options, such as send, formatting, and attachment icons, remain.}
\end{figure*}
\red{In this section, we provide the specific implementation details and user interface used in Study 2.
We developed a prototype system consisting of a Chrome extension and a backend service to enable participants to reply to emails using Gmail on a PC.
The Chrome extension detected the initiation of the reply task when participants clicked the ``Reply with AI'' button in the Gmail reply box (see Fig.~\ref{fig_UI}).
Upon clicking the button, the extension extracted the email content directly from Gmail’s DOM structure using JavaScript and sent it to a backend API endpoint implemented with FastAPI~\footnote{\url{https://fastapi.tiangolo.com}}.
The backend, hosted on an AWS EC2 instance~\footnote{\url{https://aws.amazon.com/ec2/}}, received the email content and forwarded it to the OpenAI API~\footnote{\url{https://platform.openai.com/docs/}} to generate questions or reply suggestions. 
These outputs were then returned to the Chrome extension and displayed to participant in a new reply editor.
Finally, participants revise the reply suggestions and submit them back to the Gmail reply box by clicking the ``Reply'' button.
To ensure privacy, neither the email content nor the participants' responses were accessible to the experimenters or stored on the server.}

\red{Additionally, to implement ResQ's features for generating questions and options, we provided the LLM with various contextual inputs, including the email text, subject, sender information, text from prior email interactions, and the user's details (such as name and email address). 
Furthermore, to ensure that the generated draft aligned with user expectations, the LLM was further given information outlined in Sec.~\ref{sec:generate_questions}, including the generated questions, corresponding user answers, and user preferences (\textit{e.g.}, tone, style, length, and any specific requests). 
Based on this input, the LLM produced a draft of the email reply.}

\end{document}